%% file: glasswing_paper.tex
\pdfoutput=1
\documentclass[11pt]{article}
\usepackage[margin=1in]{geometry}
\usepackage[T1]{fontenc}
\usepackage{lmodern}
\usepackage{amsmath,amssymb}
\usepackage{booktabs}
\usepackage{graphicx}
\usepackage{setspace}
\usepackage[authoryear,round]{natbib}
\usepackage[hidelinks]{hyperref}
\usepackage{xurl}
\usepackage{caption}
\usepackage[section]{placeins}
\title{Does Frontier AI Strengthen Cyber Defense?\\ Evidence from Project Glasswing\thanks{Queen Mary University of London. Mile End Road
		London E1 4NS Email: \href{mailto:t.mosk@qmul.ac.uk}{t.mosk@qmul.ac.uk}. Website: \href{https://thomasmosk.com/}{https://thomasmosk.com}.}}
\author{Thomas Mosk\\ \small Queen Mary University of London}
\date{\small November 2026}

\makeatletter
\newcommand{\econtitle}[1]{``#1\@ifnextchar.{\econ@dot}{''}}
\def\econ@dot.{.''}
\makeatother
\begin{document}
\maketitle

\begin{abstract}
\noindent Defenders and attackers race to find the same software flaws. I use the launch of Project Glasswing, which gave a small group of large US tech firms early access to Claude Mythos Preview, a model for finding software vulnerabilities, to study how access to frontier AI affects defenders, external researchers, and attackers.
Using the universe of publicly disclosed vulnerabilities, I find that firms with access more than doubled their disclosures relative to Chinese and Russian tech firms without access, disclosed about 150\% more high-impact vulnerabilities, and fixed reported flaws as quickly as before. As treated firms found more flaws themselves, external researchers, who sell their findings through bug bounties, reported about half as many vulnerabilities in treated firms' code and more in other firms' code, without leaving the market. There is no evidence that attackers exploited the additional vulnerabilities: the share exploited within a month of disclosure fell by about two thirds, mainly because a smaller share was already exploited when disclosed, while the number exploited did not change. Frontier AI helped defenders find and fix more flaws before attackers exploited them, but its restricted release creates an asymmetry in software security between firms in different countries.
\end{abstract}


\newpage
\section{Introduction}

Software vulnerabilities are a main source of cyber risk \citep{florackis2023cybersecurity,jamilov2021anatomy,ottonello2024software}. Finding them has long been slow, expensive, and skilled human work, and frontier artificial intelligence is changing that: the same models that help defenders search their code for flaws also help attackers find and exploit them. Policymakers view the most capable models as a potential shock to financial and national security. Federal Reserve Chair Powell warned the chief executives of the largest banks about the risks of Anthropic's newest model, and the Governor of the Bank of England said that it might ``crack the whole cyber risk world open.''\footnote{Reuters, ``Bessent, Powell warned bank CEOs about Anthropic model risks,'' April 13, 2026, \url{https://www.reuters.com/business/finance/bessent-powell-warn-bank-ceos-about-anthropic-model-risks-bloomberg-news-reports-2026-04-10/}; Reuters, ``Bank of England says it is testing AI risks to financial system,'' April 16, 2026, \url{https://www.reuters.com/world/uk/bank-england-says-it-is-testing-ai-risks-financial-system-2026-04-16/}.} 

A growing literature shows that generative AI raises individual productivity in tasks such as software development \citep{noy2023experimental,brynjolfsson2025generative,dellacqua2026navigating,cui2026effects}, while other work documents how AI-generated output competes with human producers \citep{reimers2026ai,goldberg2026generative,zhou2024generative}. Yet task-level productivity gains do not necessarily translate into improvements in overall production. The aggregate effects depend on the quality of AI-generated output, constraints in downstream human tasks, how producers without access to the frontier model respond, and how the resulting information affects third parties. In cybersecurity, these issues are central: a surge in AI-enabled discoveries may overwhelm human capacity to validate and remediate vulnerabilities; preferential access may allow firms to outpace independent researchers, who may consequently redirect their efforts; and additional disclosures may give attackers information about exploitable weaknesses.

This paper uses Project Glasswing to study how access to frontier AI changes such a setting. In April 2026, Anthropic gave a selected group of large US firms early access to Claude Mythos Preview, a frontier model with much stronger capabilities in vulnerability discovery and exploit development than publicly available models. Project Glasswing provided these firms with access for defensive work on their own and open-source software, while the model remained unavailable to other firms and independent researchers.

This paper makes three contributions. First, it provides evidence on how access to frontier AI affects firms' production of cybersecurity information, showing that access leads to more vulnerability disclosures without a detectable deterioration in measured quality or slowdown in remediation. Second, it examines how AI-enabled vulnerability discovery affects the allocation of independent research effort across software vendors. Third, it studies the dual-use consequences of greater vulnerability disclosure, examining whether the additional information that benefits defenders also facilitates exploitation by attackers.

The market for software vulnerabilities is well suited to studying the effects of frontier AI, because what happens to a flaw after it is found leaves a public record. Each disclosure is recorded in a common international system with a standard identifier, a publication date, the affected vendors, and measures of severity and weakness type, so the output of defenders can be compared across firms and countries. The record usually also names who found the flaw, which links a firm's disclosures to the effort of external researchers, and catalogs of exploited vulnerabilities show whether attackers later used it.

Selection into Project Glasswing was not random. Anthropic selected 13 large US firms that develop and maintain software used across the economy, giving them early access to Claude Mythos Preview for defensive security work.\footnote{Amazon Web Services (AWS), Anthropic, Apple, Broadcom, Cisco, CrowdStrike, Google, JPMorganChase, Linux Foundation, Microsoft, NVIDIA, Palo Alto Networks, and IBM.} Anthropic's regional and ownership restrictions, which predated the launch of Glasswing, excluded firms based in or majority-owned by entities in China and Russia. In addition, independent vulnerability researchers active on bug bounty platforms did not receive access. The gap in access was a gap in capability: independent evaluations found that Mythos Preview outperformed the strongest publicly available US and Chinese models in finding and exploiting vulnerabilities by a wide margin, so firms and researchers without access could use only considerably less powerful AI tools.

\begin{figure}[!t]
\centering
\includegraphics[width=\textwidth]{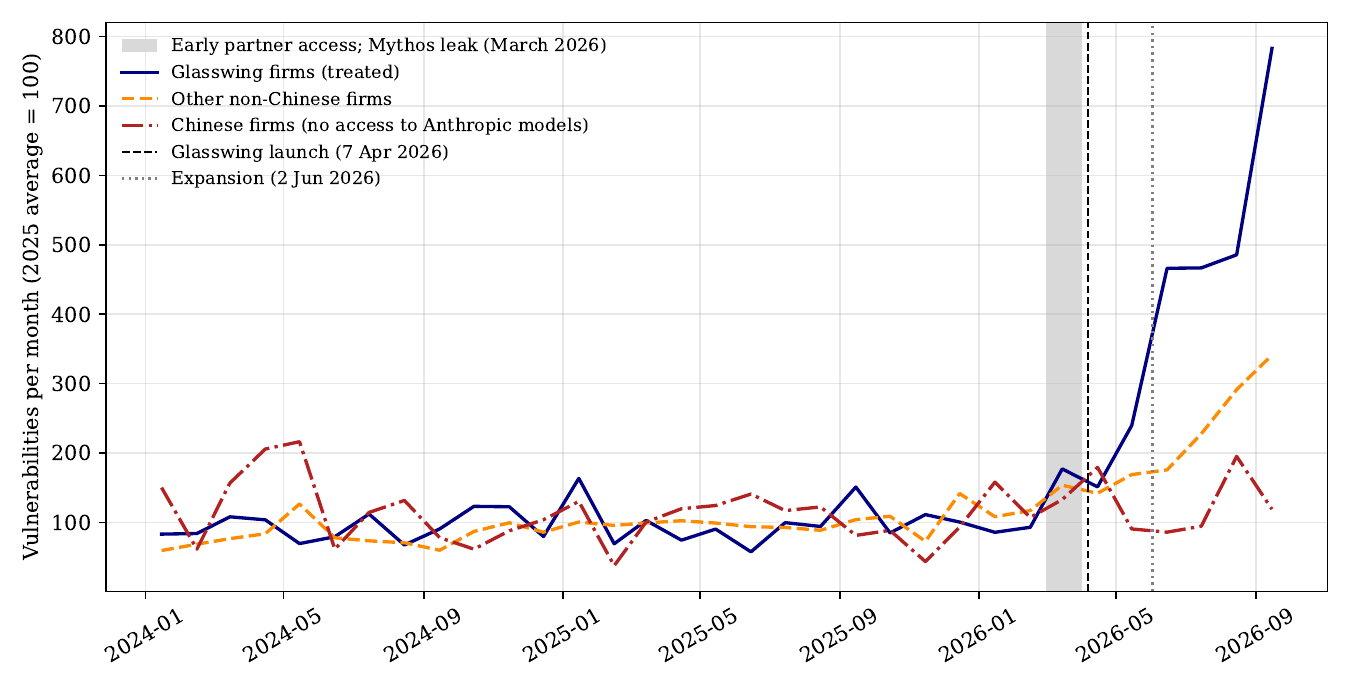}
\caption{Vulnerabilities by Group of Firms}\label{fig:cves}
\begin{minipage}{0.95\textwidth}\footnotesize \textit{Notes:} Number of vulnerabilities per month, each group indexed to its 2025 average (= 100). Glasswing firms: the twelve launch partners and IBM. Chinese firms: firms in mainland China or Hong Kong, which Anthropic's terms bar from its models. Other non-Chinese firms: all other firms. The shaded band marks March 2026, when launch partners had early access; the dashed line marks the launch (April 7, 2026) and the dotted line the expansion (June 2, 2026).\end{minipage}
\end{figure}

Figure~\ref{fig:cves} plots monthly vulnerability disclosures by Glasswing firms, Chinese firms, and all other firms, each relative to its 2025 average. The three series fluctuate around their respective 2025 averages until early 2026. Disclosures by Glasswing firms then nearly double in March 2026, when launch partners were already using Mythos, and exceed four times their 2025 average by June. This increase is consistent with external evidence. Epoch AI reports a similar post-Glasswing spike in high- and critical-severity vulnerabilities among 21 large technology firms \citep{epoch2026cve}, while Anthropic reports that its partners had identified more than 10,000 high- or critical-severity vulnerabilities within a month, most of which had not yet been disclosed \citep{anthropic2026glasswingupdate}. Chinese firms show no comparable increase. Disclosures by other non-Chinese firms also rise, most of all after the program's expansion to roughly 150 additional organizations on June 2. I therefore end comparisons with these firms in May 2026 and use Chinese firms, which were ineligible for Glasswing throughout the period, as the main comparison group.

The Chinese comparison group includes major global technology companies, such as Huawei and Lenovo. These firms have access to AI technologies for cyber defense: they could use models developed in China, including models from DeepSeek, Qwen, Z.ai, and Qihoo. However, independent evaluations indicate that the Chinese models available during the study period lagged Mythos in vulnerability discovery and exploit development. I find no evidence that Chinese disclosure changed around the launch. If China had tightened disclosure after the launch, Chinese firms would have published fewer of their vulnerabilities themselves. Researchers outside China can still publish a flaw in a Chinese product through an independent issuer, without the vendor's cooperation, so a larger share of Chinese vulnerabilities would have been published that way, and researchers, who are rewarded only when their findings are published, would have found Chinese products less attractive. Instead, Chinese firms published more of their own vulnerabilities after the launch, the share published by independent issuers did not change, and more independent researchers were credited on Chinese vulnerabilities (Appendix~\ref{sec:chinese}).

I report three main sets of results. First, access to Mythos increased firms' production of defensive information without a detectable decline in its measured quality. Relative to Chinese technology firms, firms with access more than doubled their vulnerability disclosures, an increase of about 36 per firm-month relative to the pre-launch mean. They disclosed about two and a half times as many high-impact vulnerabilities and twice as many that could be exploited remotely without credentials. This contrasts with evidence from creative markets, where generative AI has increased output while reducing average quality, both for books and AI-assisted art \citep{reimers2026ai,zhou2024generative}. I also find no evidence that the increase in disclosures slowed remediation. While productivity gains from AI coding tools can fade when work moves from writing code to shipping it, as human effort remains a bottleneck \citep{demirer2026writing}, vendors took no longer to fix reported flaws.

Second, access to the model changed the market for external vulnerability research. Large vendors buy vulnerabilities from outside researchers through bug bounty programs, which pay for each valid report; about 7,200 external persons and teams were credited with at least one vulnerability in 2024. These researchers had no access to the Mythos model, but those who searched the treated firms' code now competed with internal teams from vendors with access. A researcher who had worked only on the treated firms' code reported 57\% fewer treated-firm vulnerabilities after the launch and about two and a half times as many in other vendors' code, while the probability of being active in a month did not change. In markets for creative goods, generative AI crowds out the production of human-made content \citep{goldberg2026generative}. Human vulnerability researchers were not crowded out of the market; they moved their search to code that the model had not reached. 

Third, the increase in defensive discovery did not translate into a corresponding increase in exploitation. Vulnerability disclosures have a dual effect: they help defenders identify and fix flaws, but also provide information to attackers. Following public disclosure, attacks exploiting a vulnerability can increase by several orders of magnitude \citep{bilge2012before}. By finding vulnerabilities sooner, Mythos may shift this discovery race in defenders' favor, giving them more time to address flaws before attackers discover and exploit them. The additional disclosures did not increase attacks: the share of the treated firms' vulnerabilities exploited within one month fell by 0.9 percentage points relative to Chinese firms, about two thirds of its pre-launch rate of 1.4\%. Relative to large non-Chinese firms, the share exploited within three months fell by a similar proportion, while the number of exploited vulnerabilities did not change. The decline was concentrated in zero-day vulnerabilities, flaws that attackers were already exploiting when they were publicly disclosed. The probability of zero-day exploitation fell by about 70\% relative to Chinese firms and about 60\% relative to large non-Chinese firms. These vulnerabilities matter most because attackers can exploit them before defenders have a patch; even after an attack is detected, the flaw remains unpatched for a median of 9 days and a mean of 34 days \citep{jibunoh2026zero}. The pattern is consistent with Mythos enabling defenders to uncover vulnerabilities that attackers had not yet found.

This paper contributes to three literatures. First, it contributes to the economics of cybersecurity. Existing work measures firms' exposure to cyber risk and how it is priced \citep{florackis2023cybersecurity,jamilov2021anatomy}, and documents the costs of attacks: losses to the firms attacked \citep{kamiya2021risk}, propagation through supply chains \citep{crosignani2023pirates} and through the payment system \citep{kotidis2025propagation}, and firms' responses to breaches \citep{ademovic2025stung}. This work takes the threat as given and studies the consequences of being attacked. I study the other side of the market: how a shock to the technology of defense changes the stock of flaws available to attackers, and how attackers respond. A smaller and mostly theoretical literature on vulnerability disclosure treats the probability that a flaw is found as a parameter \citep{kannan2005market,arora2008optimal,choi2010network}, and empirical work exploits variation in disclosure regimes and markets for vulnerabilities \citep{ransbotham2012markets}. Project Glasswing moves that parameter for a known set of firms, which allows me to measure its effect on disclosure and exploitation directly.

Second, I contribute to the literature on the productivity effects of generative AI. Experiments and staggered rollouts show that AI raises the output of individual workers \citep{noy2023experimental,brynjolfsson2025generative,dellacqua2026navigating}, while the evidence on employment and earnings is mixed \citep{humlum2025large,brynjolfsson2025canaries}. In three field experiments with nearly five thousand software developers at Microsoft, Accenture, and a large electronics manufacturer, access to an AI coding assistant raised the number of completed tasks by about a quarter, mostly for junior developers \citep{cui2026effects}. In markets for digital goods, AI expands supply and crowds out human producers \citep{reimers2026ai,goldberg2026generative,zhou2024generative}, and in software the gains attenuate as work moves from writing code to shipping it, because human effort remains the bottleneck \citep{demirer2026writing}. Studies of firms measure AI by what firms choose to adopt: their AI workforce, which raises the number and success of new drug projects \citep{barwick2026drug}, their job postings \citep{li2026hiring}, or their publications and patents \citep{baviskar2026tracking}. In this paper access is allocated by the supplier of the model, not chosen by the firm, and firms barred from it form a group that never receives it. The setting also differs economically from those studied so far. Security is produced in a contest rather than as a quantity: a vulnerability is valuable to find only if attackers would otherwise have found and used it, and attackers adjust to what defenders do, so the gain from AI depends not only on how many more flaws defenders find but on whether it shifts the race between the two sides.

Third, I contribute to the literature on geoeconomic competition and restrictions on technology. Recent work studies how governments use access to critical technologies for geopolitical ends: through industrial and strategic trade policy in semiconductors \citep{liu2024sabotage,goldberg2024industrial}, through export controls, which impose costs on the firms they target \citep{crosignani2026securing} and whose effectiveness depends on international coordination and production linkages \citep{fan2026export}, and through geoeconomic pressure, to which firms respond by investing in research and development \citep{clayton2025geoeconomic}. I study a different instrument: unequal access to frontier AI. Chinese organizations have built substantial AI capacity \citep{baviskar2026tracking} and adopt generative AI quickly \citep{li2026hiring}, and the effect of AI on drug development is larger for Chinese firms than elsewhere \citep{barwick2026drug}, so the firms excluded here are not without AI. Project Glasswing gave a frontier model to a group of mostly US firms, while the provider's terms exclude firms controlled from China and Russia. The restriction is private rather than governmental, but, like an export control, it allocates a strategic technology along geopolitical lines. Relative to the excluded firms, firms with access roughly doubled the latent flaws they removed from their software. Competition over frontier AI therefore concerns not only who can build the technology but also who can use it to protect critical digital infrastructure.

My study has some limitations. I observe disclosed vulnerabilities rather than the underlying stock of flaws or the total number discovered, and disclosure counts reflect discovery, validation, and the decision to publish. Because Anthropic chose the firms with access and the main comparison group consists of Chinese firms operating under different disclosure rules, the estimates measure access to Mythos Preview relative to the AI tools available to these firms, not the effect of AI relative to no AI; the comparison with large non-Chinese firms provides a check but ends before the program expanded in June 2026. Finally, observing vendor fixes does not show whether or when users installed the patches, so the results concern disclosure and vendor remediation rather than the security of software in use.

The remainder of the paper is organized as follows. Section~\ref{sec:background} explains the institutional background,
Section~\ref{sec:data} describes the data and sample construction, Section~\ref{sec:firm} presents the firm-level evidence, Section~\ref{sec:researchers} the researcher-level evidence, Section~\ref{sec:attack} the response of attackers, and.
Section~\ref{sec:con} concludes.

\section{Institutional Background}\label{sec:background}

\subsection{Project Glasswing}\label{sec:glasswing}

Project Glasswing gave a selected group of firms early access to Claude Mythos Preview, a model that finds and exploits software vulnerabilities with little human help, before comparable capabilities were available to other firms and researchers.

\paragraph{Claude Mythos Preview.} Claude Mythos Preview is a general-purpose language model that finds and exploits software flaws far better than earlier models. Anthropic did not train it for this purpose; the capability emerged from better reasoning about code and greater autonomy \citep{anthropic2026mythosred}. Given a short instruction and access to a program, the model reads the code, forms hypotheses about where it might fail, tests them, and returns a report with a working exploit, with little human guidance. What used to require weeks of work by a skilled researcher can therefore be done at scale across large codebases, so the model sharply lowers the cost of finding vulnerabilities. Independent evaluations confirm that the gap is large. The UK AI Security Institute found that Mythos Preview solved far more expert-level hacking exercises than earlier models and was the first model to carry out a simulated attack on a corporate network from start to finish, and evaluations by the US government and by academic researchers rank it well ahead of the strongest public models of Anthropic and other developers.\footnote{Mythos Preview solved 73\% of the institute's expert-level capture-the-flag exercises, which no model could solve a year earlier. In the 32-step simulated attack, from reconnaissance to full takeover, it completed all steps in 3 of 10 attempts and 22 steps on average, against 16 for Claude Opus 4.6; the test networks had no active defenders \citep{aisi2026mythos}.}

\paragraph{Access and selection.} Anthropic announced Project Glasswing on April 7, 2026, and chose which organizations received access. Twelve launch partners (Amazon Web Services, Anthropic, Apple, Broadcom, Cisco, CrowdStrike, Google, JPMorganChase, the Linux Foundation, Microsoft, NVIDIA, and Palo Alto Networks) received the model for defensive work on their own and open-source software. Anthropic committed up to \$100 million in usage credits and \$4 million in donations to open-source security organizations, and stated that it would not make the model generally available \citep{anthropic2026glasswing}. The launch partners were not selected at random. All are headquartered in the United States, and they are among the largest producers of the operating systems, browsers, cloud services, network equipment, and security software on which other firms depend.\footnote{At the launch, Anthropic also ``extended access to over 40 additional organizations that build or maintain critical software infrastructure,'' without naming them \citep{anthropic2026glasswing}. None of these organizations can be Chinese or Russian, because Anthropic's terms bar companies controlled from China and Russia from all of its models.} IBM announced on May 19, 2026, that it had joined \citep{ibm2026glasswing}, and on June 2, 2026, Anthropic extended access to roughly 150 further organizations in more than 15 countries, whose names were not disclosed. 

\paragraph{Excluded firms and researchers.} Firms in China and Russia, such as Huawei, Lenovo, Yandex, and Kaspersky, could not obtain the model at any point, and their exclusion predates the model. Neither country is among the regions in which Anthropic offers its services \citep{anthropic2026supported}. Russian firms were already unable to use Anthropic's commercial services, and in September 2025, seven months before the launch, Anthropic extended the restriction to companies majority-owned by entities in unsupported regions such as China, wherever they operate \citep{anthropic2025restrictions}. 
External researchers are people and teams outside the vendor, mostly independent individuals, who search other companies' software for flaws. They report what they find to the vendor, often through a bug bounty program that pays per valid report, or through a coordinator such as the Zero Day Initiative. In return they are credited in the vendor's advisory and paid a bounty. External researchers did not receive access to Mythos and could use only public models.

\paragraph{Other AI models.} US and Chinese developers released new AI models throughout the sample period, but independent evaluations show that Claude Mythos Preview had substantially better cyber capabilities than any of them. The most informative evaluations come from two government bodies that test models independently of their developers, the US Center for AI Standards and Innovation (CAISI) and the UK AI Security Institute. One of CAISI's tests, ExploitBench, asks a model to develop working exploits for software flaws in controlled environments and reports the share of attack goals it achieves. On this test Mythos Preview scores 57.2, against 38.1 for Claude Opus 4.8, Anthropic's most capable public model at the time,\footnote{Anthropic released Claude Opus 4.6 on February 5, 2026, Claude Opus 4.7 on April 16, 2026, and Claude Opus 4.8 on May 28, 2026.} and 40.7 for GPT-5.5, the strongest model of another US developer, OpenAI \citep{nist2026glm52,openai2026gpt55}.\footnote{An academic benchmark built from 898 real-world vulnerabilities in user programs, the V8 JavaScript engine, and the Linux kernel gives the same ranking: Claude Mythos Preview produced working exploits for 157 of them and GPT-5.5 for 120 \citep{wang2026exploitgym}.} The public versions of these models also restrict high-risk cyber uses, and both developers offer less restricted versions only to vetted security professionals \citep{anthropic2026glasswingupdate}. Chinese firms, which could use none of these US models, relied on domestic models that were further behind. The most capable Chinese models are DeepSeek V4 and Z.ai's GLM-5.2.\footnote{DeepSeek released V4 on April 24, 2026, and Z.ai released GLM-5.2 on June 16, 2026 \citep{deepseek2026v4,nist2026glm52}.} GLM-5.2 scores 21.4 on ExploitBench, less than half the score of Mythos Preview \citep{nist2026glm52}, and the UK AI Security Institute finds that both Chinese models perform on cyber tasks like US models released four to seven months earlier \citep{ukaisi2026openweight}.

\subsection{Vulnerabilities and Disclosure Policies}\label{sec:disclosure}

\paragraph{How vulnerabilities are disclosed.} A researcher who finds a flaw reports it to the vendor, directly, through a bug bounty program or through a coordinator such as the Zero Day Initiative or the CERT Coordination Center, and commonly allows the vendor a fixed period, often 90 days, to prepare a fix before details become public. A Common Vulnerabilities and Exposures (CVE) identifier is assigned by an organization authorized by the CVE Program, which I call an issuer. Large vendors such as Microsoft, Apple, Google, and Cisco, and Chinese vendors such as Huawei and ZTE, are their own issuers; independent issuers such as VulDB, Patchstack, and MITRE issue identifiers for flaws reported by researchers. When a vendor is not an issuer itself or does not respond, a researcher can obtain an identifier from an independent issuer and publish the vulnerability without the vendor's cooperation. An issuer first reserves an identifier, typically when it accepts a report or confirms a flaw, and publishes the CVE record once the vulnerability is disclosed, usually together with a fix and a security advisory. Vendors often disclose on fixed release cycles, such as Microsoft's monthly Patch Tuesday and Oracle's quarterly Critical Patch Updates. The official CVE List, maintained by the CVE Program, holds the authoritative records, and an issuer can later reject a record that turns out to be invalid or a duplicate. The National Vulnerability Database (NVD), operated by the US National Institute of Standards and Technology, adds severity scores, weakness types, and affected-product configurations.

\paragraph{Disclosure in China.} China regulates disclosure more tightly.\footnote{Russia keeps its own national database, maintained by the Federal Service for Technical and Export Control, and few Russian vendors publish their vulnerabilities as CVEs. Russian firms form an additional comparison group (Appendix~\ref{sec:construction}).} Since September 1, 2021, its Regulations on the Management of Security Vulnerabilities of Network Products require vendors to report vulnerabilities in their products to the Ministry of Industry and Information Technology within two days, prohibit researchers from publishing details before a fix is available, and bar sharing vulnerability details with overseas organizations other than the vendor.\footnote{The Record, ``Chinese government lays out new vulnerability disclosure rules,'' July 2021, \url{https://therecord.media/chinese-government-lays-out-new-vulnerability-disclosure-rules}.} A tighter policy after the launch would bias the comparison, because the disclosures of Chinese firms would then fall for reasons unrelated to AI. 
I find no evidence that China changed its disclosure policy around the launch of Claude Mythos Preview. The 2021 regulations were not amended, and the data show none of the signs that a tighter policy would leave (Appendix~\ref{sec:chinese}): Chinese vulnerabilities did not fall in any disclosure channel, the share published by independent issuers outside the vendors' control did not change, and more independent researchers were credited on Chinese vulnerabilities after the launch, including on the vulnerabilities that Chinese vendors publish themselves.

\paragraph{What the comparison requires.} Despite these rules, the vulnerabilities of Chinese firms are recorded largely in the same system as those of US firms. Of the 2,099 vulnerabilities of Chinese firms published between October 2025 and September 2026, 94\% have a CVE identifier and appear in the National Vulnerability Database; the remaining 122 appear only in the China National Vulnerability Database. Excluding these entries, which also removes the Chinese firms that appear only in that database, does not change the main findings of the paper. The Chinese rules do shape the level and channels of disclosure: Chinese firms disclose fewer vulnerabilities, 44\% of their CVEs are issued by VulDB on behalf of researchers rather than by the firms themselves, and entries without a CVE are published in irregular batches. The design does not require Chinese and US firms to disclose in the same way. It requires that the proportional difference between their disclosures would have stayed constant without Project Glasswing. The Chinese rules have been in force since September 2021, four and a half years before the launch, and I am not aware of changes to them around April 2026; the event study in Section~\ref{sec:firm} shows no differential trend in the months before access, and a placebo launch a year earlier finds no effect.

\section{Data}\label{sec:data}

\subsection{Sample construction}\label{sec:sample}

The unit of observation is a publicly disclosed software vulnerability. A vulnerability that a vendor or a researcher reports through the industry's common disclosure system receives a Common Vulnerabilities and Exposures (CVE) identifier. The official CVE List, maintained by the CVE Program, records for every CVE the organization that issued the identifier, the date on which the identifier was reserved, the date on which the vulnerability was published, the affected vendors and products, links to advisories and fixes, and the parties credited with finding the flaw. I combine the CVE List as of October 28, 2026, with the National Vulnerability Database of the US National Institute of Standards and Technology, which adds the affected product configurations, the type of weakness, and a severity score. Because Chinese vendors can also disclose vulnerabilities only domestically, I add the China National Vulnerability Database, which records vulnerabilities in Chinese products with or without a CVE identifier. Most of its entries for Chinese firms duplicate a CVE; the 720 entries without one enter the sample as separate vulnerabilities. The sample contains all 176,384 CVEs published between January 1, 2024, and October 31, 2026, and these 720 entries, 177,104 vulnerabilities in total. Appendix~\ref{sec:construction} describes each source and each construction step.

I attribute each CVE to the firms whose products it affects. A CVE belongs to a firm if the firm is named as the vendor of an affected product, if the firm issued the identifier for its own product, or if the CVE links to the firm's own security advisory. Security research teams that publish flaws in other firms' products, such as Cisco Talos or Google Project Zero, therefore do not attribute those flaws to their parent firm. Product databases also list the operating system on which an affected application runs, so that a flaw in the Chrome browser is listed with Windows and macOS. I attribute such a CVE only to the vendor of the application. This rule removes 3,036 CVEs from Microsoft and 2,244 from Apple over the sample period, most of them flaws in software of Google, Adobe, and IBM. A CVE can belong to more than one firm.

\paragraph{Treatment at the firm level.} The treated firms, which I call Glasswing firms, are the twelve launch partners of Project Glasswing and IBM; Broadcom includes VMware. The treatment starts in April 2026 for the launch partners and in May 2026 for IBM. The main comparison group consists of Chinese firms. Since September 2025 Anthropic has barred companies that are majority-owned or controlled by Chinese entities from its models, so no Chinese firm could join Project Glasswing at the launch or after the expansion. I identify 27 Chinese technology firms by name and treat every other Chinese vendor in the data as its own firm, which gives 267 Chinese firms with at least one vulnerability between January 2024 and October 2026. Their vulnerabilities include their CVEs and their entries in the China National Vulnerability Database without a CVE. A second comparison group, used in robustness tests and in the analysis of attackers, consists of 26 large non-Chinese firms that were not Glasswing members at the launch: nine cybersecurity vendors, six software and platform firms, six semiconductor and networking firms, and five large banks. Some of them may have gained access after the expansion of June 2, 2026, so comparisons with them end in May 2026. A third comparison group, also used in robustness tests, adds Russian firms, which Anthropic's terms bar from its models as well. Few Russian vendors publish their vulnerabilities as CVEs: ten have at least one CVE between January 2024 and October 2026, and nine of them have one in the analysis sample (Appendix~\ref{sec:construction}). Appendix Table~\ref{tab:firms} lists all firms. In the vulnerability-level sample, a vulnerability belongs to the Glasswing group if it affects a Glasswing firm, to the Chinese group if it affects a Chinese firm, and to the group of other non-Chinese firms otherwise. Of the 176,384 CVEs, 21,361 (12.1\%) affect a Glasswing firm.

\paragraph{Exploitation.} I date the first exploitation of a vulnerability by attackers with two catalogs of vulnerabilities known to be exploited in the wild, kept by the US Cybersecurity and Infrastructure Security Agency (CISA) and by the security firm VulnCheck. The second catalog is larger and records the date of the first public report of exploitation. Together they identify exploitation for 5,718 CVEs. I date the first public exploit code with four collections of published exploits, which together cover 16,801 CVEs. I also collect the daily scores of the Exploit Prediction Scoring System on the last day of every month from April 2021 to October 2026 (Section~\ref{sec:attack}). Both dates record when exploitation or exploit code became publicly known, not when it began.

\subsection{Researcher data}\label{sec:hunterdata}

I use the term \emph{researcher} for every person or team credited with finding a vulnerability, whether it works for the affected vendor or not. Researchers employed by the affected vendor form its internal team.

I identify the researchers who found each vulnerability by combining several sources, because the official record names a finder for fewer than half of the CVEs. Vendors thank the researchers who reported a flaw in their own security advisories, bug bounty and disclosure programs name the researchers they paid, open-source projects credit reporters in their advisories and in the code change that fixes a flaw, and researchers often publish write-ups to which the CVE record links. I collect these acknowledgments from the CVE List, the advisories of 17 vendors (including the Glasswing firms Microsoft, Google, Apple, Cisco, and NVIDIA), two disclosure programs, and three databases of open-source advisories (Appendix Table~\ref{tab:sources}). After matching across these sources, I observe at least one finder for 73.0\% of the CVEs in the sample (128,754). The remaining 47,630 CVEs stay in the sample with an unknown finder.

A credit names one or more people, teams, or tools, often with an affiliation, as in ``A, and B of Company X, working with Program Y.'' I split each credit into the entities it names, separate each name from its affiliation and from intermediaries such as bounty platforms, and classify each entity as a person, a team or organization, an AI system, an automated tool, or an anonymous researcher. Each entity receives a permanent identifier based on its standardized name. Because the same person is often credited under a real name in one source and an online handle in another, I merge identifiers that share a handle or an account on GitHub, the platform on which most open-source code is hosted. A researcher's affiliation is the organization with which it is most often credited, and organizations are matched to their parent firms, so that Google Project Zero belongs to Google and Talos to Cisco. The resulting database contains 41,299 researchers: 39,597 persons, 1,649 teams or organizations, 14 AI systems, 10 automated tools, 28 reporters known only by a pseudonym, and one pooled anonymous researcher.

\paragraph{Treatment at the researcher level.} Project Glasswing reaches researchers in two ways. Researchers who work for a launch partner have direct access to the model, as do those who work for IBM from May 2026. Researchers outside the Glasswing firms have no access, but those who search the Glasswing firms' code now compete with the firms' AI-assisted search for the same flaws. Section~\ref{sec:researchers} measures this exposure by the share of a researcher's vulnerabilities in 2024 that affected Glasswing firms. For the descriptive statistics in Table~\ref{tab:t1}, a researcher is treated if it reported at least one vulnerability of a Glasswing firm between January 2024 and March 2026, which holds for 2,783 of the 18,650 researchers credited before the launch (14.9\%).

\subsection{Measures}\label{sec:measures}

\paragraph{Firm level.} The main outcome is the number of vulnerabilities attributed to a firm in a month, in total and by type; Section~\ref{sec:firm} defines the types. The \emph{disclosure gap} of a CVE is the number of days between the reservation of its identifier and its publication. A vendor reserves an identifier once it has accepted a report or confirmed a flaw, so the gap measures the period in which the vendor knows of the vulnerability and the public does not; it is a lower bound on the time from discovery to disclosure.

\paragraph{CVE level.} \emph{Treated} equals one if the CVE affects a Glasswing firm, and \emph{Post} equals one for CVEs published from April 2026. \emph{Exploited within two (three) months} equals one if the first evidence of exploitation in the wild comes no later than 60 (90) days after publication. It is defined only for CVEs that had been public for the full window by October 28, 2026, so that recent CVEs are not mechanically less likely to be exploited.

The \emph{finder type} has four mutually exclusive categories. A CVE is found by the vendor's \emph{internal team} if a credited researcher works for the affected vendor or the vendor states that the flaw was found internally. Otherwise it is found by \emph{another organization or project} if a finder is credited as a team or organization, as a person with an affiliation (a security firm, a university, an open-source project, or an AI laboratory), or as an AI system or automated tool. Otherwise it is found by an \emph{individual researcher} if a person is credited without an affiliation or anonymously. If no source names a finder, the finder is \emph{unknown}. In the main analysis a CVE counts as an internal find only on explicit evidence. Two kinds of CVEs are probably internal finds without an explicit statement: CVEs that the vendor's own advisory lists without crediting anyone, and Linux kernel fixes whose code change credits no reporter. Appendix Table~\ref{tab:inferred} shows the finder types when these CVEs count as internal.

\paragraph{AI exposure.} A CVE is \emph{AI-exposed} if its weakness type is a memory-safety flaw, such as reading or writing beyond the bounds of a buffer or using memory after it has been released (Appendix Table~\ref{tab:adcsum} lists the types). The measure follows the literature that classifies occupations by how exposed their tasks are to AI \citep{eloundou2024gpts}, applied to types of flaws. Memory-safety flaws are the class that AI finds most easily, because they make a program crash or trigger an automatic memory checker, so a model can propose a candidate flaw and verify it at almost no cost. Most documented discoveries by AI systems are of this kind, and AI developers focus their searches on them for this reason.\footnote{Examples include the memory flaws found by Google's Big Sleep agent in the SQLite database engine \citep{bigsleep2024} and by AI-written test programs in the OpenSSL encryption library \citep{google2024ossfuzz}, and the CyberGym benchmark, which scores AI agents by whether they trigger a memory-checker crash \citep{wang2025cybergym}. Anthropic's researchers write: ``We focused on searching for memory corruption vulnerabilities, because they can be validated with relative ease'' \citep{anthropic2026zerodays}. Models that only read code, without running it, detect vulnerabilities poorly \citep{ullah2024llms,ding2025vulnerability}.} Of the CVEs published between January 2024 and September 2026, 17.3\% are AI-exposed: 36.3\% of those of Glasswing firms and 19.8\% of those of the 26 non-Chinese comparison firms.

\paragraph{Types of vulnerabilities.} Beyond AI exposure, I classify vulnerabilities by what they are worth to defenders and attackers. \emph{Machine-verifiable} flaws are those that a crash, a memory checker, or a crafted input can confirm; besides memory-safety flaws they include the injection of database commands or scripts through user input, which automated scanners have found for many years. \emph{Complex weaknesses} are their counterpart: flaws whose presence no automatic test can confirm, such as missing authorization checks, broken authentication, weak cryptography, and race conditions, which require an understanding of how the software is meant to behave. \emph{High-impact} vulnerabilities have an impact component of the standard severity score, which rates the damage to the confidentiality, integrity, and availability of a system, above the median of all CVEs published between January 2024 and October 2026. \emph{Remote, no credentials} vulnerabilities can be exploited over the network without an account and without the victim's help, the flaws that matter most for attacks at scale. \emph{Many releases} marks vulnerabilities present in more product versions than the median CVE; because the date on which a flaw entered the code is not observed for closed-source products, the number of affected releases serves as a proxy for how long the flaw survived in shipped code.

\paragraph{Researcher level.} \emph{Vulnerabilities} counts the CVEs credited to a researcher in a month, each weighted by $1/N$ when $N$ finders are credited on the CVE, so that a CVE contributes one in total. I split this count into vulnerabilities at treated and at non-treated firms, and count the AI-exposed vulnerabilities. \emph{Active} equals one in months with at least one credited CVE. \emph{Affiliated with a treated firm} equals one if the researcher's affiliation is a Glasswing firm. \emph{Team or organization} marks researchers credited as a team, and \emph{unaffiliated individual} marks persons credited without an affiliation.

\subsection{Descriptive statistics}\label{sec:desc}

Table~\ref{tab:t1} reports descriptive statistics. Panel A describes the 177,104 vulnerabilities: 176,384 CVEs and 720 entries of Chinese firms in the China National Vulnerability Database without a CVE. Of these, 12.1\% affect a Glasswing firm, and about 1\% are exploited within two or three months of publication. Of the CVEs with a known weakness type, 18.7\% are AI-exposed. An individual researcher found 52.6\%, another organization or project 16.5\% and the vendor's internal team 4.0\%; for 27.0\% the finder is unknown. Before the launch, the CVEs of treated firms differ from other CVEs in three ways. They are about twice as likely to be exploited within two months (1.8\% against 0.9\%), consistent with the concentration of attacks on widely deployed products. They are more than twice as often AI-exposed (33.0\% against 15.5\%). The treated firms make operating systems, browsers, and network software written largely in languages without automatic memory management, while a third of the other CVEs before the launch come from three issuers that specialize in plugins for the WordPress website software and in small web applications, whose flaws are mostly injection flaws. And the CVEs of treated firms are found more often by the vendors' own teams (13.3\% against 1.5\%) and by other organizations (27.2\% against 16.8\%) and less often by individual researchers (34.4\% against 56.1\%), while their finder is unknown about as often (25.1\% against 25.6\%). The difference-in-differences design compares changes within these groups, so fixed effects absorb the level differences, but the different mix of finders matters for how AI affects researchers.

Panel B describes the 18,650 researchers credited before the launch, observed over 34 months. A researcher is credited with 0.15 vulnerabilities in an average month and is active in 6.1\% of months; the distribution is very skewed, with a standard deviation of 2.1. Before the launch, treated and non-treated researchers were equally productive (0.143 and 0.146 vulnerabilities per month), but treated researchers were more often active (9.0\% against 6.0\% of months). They split their output between treated firms (0.082) and other firms (0.061), while non-treated researchers by construction work only on other firms. Treated researchers find more AI-exposed vulnerabilities (0.048 against 0.016 per month), are more often affiliated with a treated firm (14.7\% against 1.3\% of researcher-months) and are less often unaffiliated individuals (41.0\% against 75.7\%). Researcher fixed effects absorb these level differences.

The rise in Glasswing disclosures begins in March 2026, a month before the launch, and the timing points to early access rather than a pre-trend, because the launch announcement states: ``In addition to our own work, many of our partners have already been using Claude Mythos Preview for several weeks'' \citep{anthropic2026glasswing}. Because March is partly treated, I use February 2026 as the reference month in the event studies.

\input{tables/table1_sample.tex}

\section{Firm-Level Evidence}\label{sec:firm}

This section asks whether Project Glasswing changed the vulnerabilities that the Glasswing firms disclosed, and whether the change amounts to better security or only to more reports.

\paragraph{Specification.} I compare the Glasswing firms, the twelve launch partners and IBM, with Chinese firms. Since September 2025, Anthropic has barred companies that are majority-owned or controlled by Chinese entities from using its models, so no Chinese firm could join Project Glasswing at the launch or in the June 2026 expansion. Chinese firms are therefore a never-treated group for the whole sample period, unlike the other non-Chinese firms, many of which may have gained access after June 2, 2026. They also make similar products: operating systems, phones, routers, cameras, and network equipment, much of it written in languages without automatic memory management, and before the launch 41\% of their vulnerabilities were AI-exposed, against 33\% at the Glasswing firms (Table~\ref{tab:t1}). The Chinese group consists of 27 Chinese technology firms identified by name, from Huawei, Lenovo, ZTE, Xiaomi, Tencent, and Alibaba to the router makers Tenda and TOTOLINK, and the other Chinese vendors in the data, each as its own firm. The firms with the most vulnerabilities in the window are Tenda (526), TOTOLINK (380), Huawei (189), TP-Link (173) and Foxit (97), which together account for about two thirds of the Chinese firms' vulnerabilities. Their vulnerabilities include the CVEs attributed to them and their entries in the China National Vulnerability Database without a CVE (Section~\ref{sec:sample}).

The sample is a balanced firm-month panel for a symmetric window of six months before and six months after the launch, October 2025 to September 2026, for the 13 Glasswing firms and the 267 Chinese firms with at least one vulnerability between January 2024 and October 2026. I estimate
\begin{equation}
\log\left(1+N^{k}_{it}\right) = \alpha_i + \delta_t + \beta\,\text{Glasswing}_i\times\text{Post}_t + \varepsilon_{it},
\label{eq:firm}
\end{equation}
where $N^{k}_{it}$ is the number of vulnerabilities of type $k$ of firm $i$ published in month $t$, $\alpha_i$ and $\delta_t$ are firm and year-month fixed effects, and $\text{Glasswing}_i\times\text{Post}_t$ equals one from the firm's own access month, April 2026 for the launch partners and May 2026 for IBM. Because the panel is balanced and firm-months without vulnerabilities enter with $N^k_{it}=0$, every regression uses the same 3,360 firm-months, and the estimates do not condition on which vulnerabilities were disclosed. A change in the composition of disclosures appears as a difference between the coefficients for different types. Equation~\eqref{eq:firm} identifies the effect of access if, without Project Glasswing, the Glasswing firms' disclosures would have grown at the same proportional rate as those of the Chinese firms. The logarithmic outcome makes this a statement about proportional growth, which is the natural assumption when a Glasswing firm discloses five times as many vulnerabilities as a large Chinese firm.

\paragraph{Inference.} Standard errors are clustered by firm. With only 13 Glasswing firms, however, the clustered standard error may understate the uncertainty, because the variation of the treated firms is estimated from few clusters and the Glasswing firms are much larger, and therefore noisier, than most Chinese firms \citep{ferman2019inference}. I therefore also report the $p$-value of \citet{ibragimov2010tstatistic,ibragimov2016inference}, which estimates the effect separately for each Glasswing firm against all Chinese firms and tests the mean of the 13 estimates with a $t$-test with 12 degrees of freedom. This test uses only the variation across the treated firms themselves, so it does not assume that Glasswing and Chinese firms are equally noisy, and it allows the effect to differ across firms.

\begin{figure}[!htbp]
\centering
\includegraphics[width=0.9\textwidth]{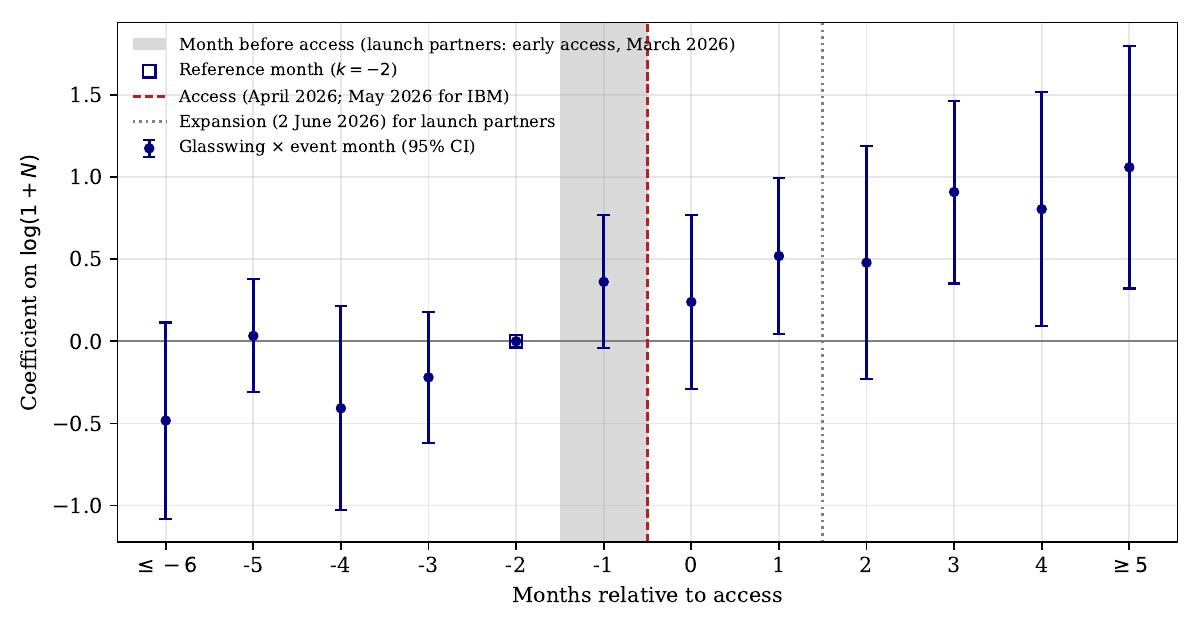}
\caption{Dynamic Difference-in-Differences: Glasswing Firms and Chinese Firms}\label{fig:firmes}
\begin{minipage}{0.92\textwidth}\footnotesize \textit{Notes:} Coefficients on Glasswing interacted with indicators for the month relative to the firm's access month (April 2026 for the twelve launch partners, May 2026 for IBM), from a regression of $\log(1+N)$ on these interactions and firm and year-month fixed effects, for the 13 Glasswing firms and 267 Chinese firms, October 2025 to September 2026. $N$ is the number of vulnerabilities of the firm published in the month. The reference is two months before access (open square), February 2026 for the launch partners, the last month before they had early access; the end points bin six or more months before and five or more months after access. Bars show 95\% confidence intervals based on standard errors clustered by firm. The shaded band marks the month before access (March 2026 for the launch partners: early access and the leak of the model); the dashed line marks access and the dotted line the expansion on June 2, 2026, for the launch partners.\end{minipage}
\end{figure}

\paragraph{Event study.} Figure~\ref{fig:firmes} plots the coefficients of a dynamic version of equation~\eqref{eq:firm} for all vulnerabilities, in months relative to each firm's access month. The reference is two months before access, February 2026 for the launch partners, the last month before they had early access. The coefficients for three to six months before access are between $-0.48$ and $0.03$, individually insignificant and jointly insignificant ($p=0.19$). The coefficient turns positive in the month before access, March 2026 for the launch partners, when they had early access (0.36), stays positive in the first two months of access (0.24 and 0.52) and rises to between 0.48 and 1.06 thereafter, after the expansion.

\paragraph{Main estimates.} Table~\ref{tab:firm} reports the baseline estimate and the tests of the identifying assumption. Column (1) shows that the Glasswing firms' disclosures rose by 0.784 log points relative to the Chinese firms, an increase of about 120\%. At the Glasswing firms' pre-launch mean of 29 vulnerabilities per firm and month, the estimate corresponds to about 36 additional disclosures per firm and month. The Ibragimov-M\"uller $p$-value is 0.002. Other methods designed for few treated groups, including the wild cluster bootstrap and the heteroskedasticity-corrected bootstrap of \citet{ferman2019inference}, give $p$-values below 0.005 (Appendix Table~\ref{tab:inference}). Column (2) estimates the same comparison by Poisson pseudo-maximum likelihood on the number of vulnerabilities. Column (3) excludes Anthropic, the provider of the model and itself a launch partner. Leaving out any other Glasswing firm one at a time gives estimates between 0.691, without Google, and 0.872, without the Linux Foundation, and are all significant at the 1\% level (Appendix Table~\ref{tab:firmloo}). Because Chinese firms were barred from all of Anthropic's models from September 2025, the comparison could also reflect access to its generally available models, such as Claude Opus 4.6, released in February 2026. Appendix Table~\ref{tab:firmopus} adds an indicator for the Glasswing firms from February 2026 to columns (1) to (4). The effect of access to Claude Mythos Preview remains. 

One concern is that the largest vendors adopted other AI tools for vulnerability research earlier than Chinese firms, so that their disclosures would have diverged gradually even without the launch. The event study in Figure~\ref{fig:firmes} shows no such divergence in the months before access. Column (4) extends the window to January 2024 and allows the Glasswing firms their own linear trend. Column (5) dates a placebo launch one year before access, in the window from October 2024 to September 2025, and finds no effect.

Columns (6) to (8) change the comparison group. Column (6) compares the Glasswing firms with the 26 large non-Chinese firms from October 2025 to May 2026, before any of them could have gained access through the expansion. Column (7) adds to the Chinese firms the Russian firms, which Anthropic's terms also bar from its models. Because few Russian vendors publish their vulnerabilities as CVEs, the nine Russian firms add only 10 vulnerabilities to the window and leave the estimate unchanged. Column (8) uses all four groups in one regression and treats the 26 non-Chinese firms from the expansion of June 2026, with the Chinese and Russian firms as the never-treated group. 

\input{tables/table2_identification.tex}

\paragraph{Types of vulnerabilities.} The increase in disclosures raises two questions. The first is whether the model found only the flaws that AI is known to find easily, memory-safety flaws that a crash confirms, or also flaws that require an understanding of how the software is meant to behave. The second is whether the additional vulnerabilities matter. No single measure captures the value of a vulnerability, but the standard classifications indicate how much damage a flaw allows, how easily it can be exploited at scale, and how long it survived in shipped code. Table~\ref{tab:firmtype} estimates equation~\eqref{eq:firm} separately for each type, defined in Section~\ref{sec:measures}.

The increase is not confined to the flaws that AI finds most easily. AI-exposed vulnerabilities roughly doubled, but other vulnerabilities rose at least as much, and complex weaknesses, which no automatic test can confirm, rose about as much as machine-verifiable ones. Nor did the additional vulnerabilities consist of minor flaws. The rise was largest for high-impact vulnerabilities, which increased about two and a half times, and vulnerabilities that can be exploited remotely without credentials and those present in many releases rose broadly in line with all disclosures. At the pre-launch means, a Glasswing firm disclosed about 23 additional high-impact vulnerabilities a month. Access to the model therefore raised the disclosure of every type of flaw, including the severe, remotely exploitable, and long-lived flaws that matter most for security.

\input{tables/table3_firm_types.tex}

\paragraph{New discoveries or a backlog?} The increase could reflect new discoveries or the release of a backlog of flaws that the Glasswing firms had known about but not yet published. The two explanations differ in the disclosure gap, the time between the reservation of an identifier and publication: new discoveries are published soon after reservation, while a released backlog consists of flaws that have waited a long time. The data favor new discoveries. Relative to Chinese firms, disclosures published within three months of reservation rose sharply, while those that had waited longer did not change (Appendix Table~\ref{tab:firmgap}), and the median gap of Glasswing vulnerabilities fell from 75 days before access to 22 days after it. A vendor that finds a flaw itself may reserve the identifier only when the fix is ready, so a short gap does not prove a short time from discovery to disclosure. Under either reading, however, the additional vulnerabilities were not old reports released at once.

\paragraph{Who found the additional vulnerabilities?} If access to the model produced new information inside the firms, the additional disclosures should come from the firms' own security teams; if it changed only how the firms handled reports by others, they should come from external researchers. Both rose (Appendix Table~\ref{tab:firmfinder}), but internal finds rose far more. The average Glasswing firm published about 4 internally found vulnerabilities a month before access and 32 after, an eightfold increase, while external reports roughly tripled. Internal finds account for about a third of the rise in disclosures, and for about two fifths when the vulnerabilities that a vendor's own advisory lists without any credit are counted as internal. The internal increase is the most direct evidence that access to the model produced new information inside the firms. The external increase came mostly from researchers who had not been credited in 2024: Glasswing vulnerabilities found only by such newcomers quadrupled, while those with a finder credited in 2024 roughly doubled. The 2024 researchers followed in Section~\ref{sec:researchers} therefore account for a shrinking share of the external reports to the Glasswing firms.

\paragraph{Remediation: was there a human bottleneck?} If AI raised discovery while humans still had to assess, fix, test, and ship every flaw, remediation could have become the bottleneck, as in software production more generally \citep{demirer2026writing}. I measure remediation at two stages. At the vendor stage, the Zero Day Initiative records the date on which it reported each flaw to the vendor and publishes its advisory once the vendor has shipped a fix, so the time between the two measures the vendor's remediation time for a stream of reports that does not depend on who found other flaws. Glasswing vendors did not become slower at fixing these reports relative to other vendors; if anything they became faster, and the estimates rule out a slowdown of more than about 13\% (Table~\ref{tab:patching}). At the downstream stage, Linux distributions such as Debian, Ubuntu, Red Hat, and SUSE, whose security teams had no access to the model, shipped fixes for Glasswing vulnerabilities as quickly as before: the median time to a fix stayed at one day, and the estimates rule out a decline of more than about 11 percentage points in the share fixed within 90 days. Among the flaws that were disclosed, the doubling of disclosures therefore did not congest either the vendors or the distributions that pass fixes to users. These measures cover only flaws that were fixed and disclosed, however. Anthropic reports that its partners had found more than 10,000 high- or critical-severity vulnerabilities within a month, most of them not yet disclosed \citep{anthropic2026glasswingupdate}, far more than the roughly 470 additional disclosures a month implied by Table~\ref{tab:firm}. Fixing capacity may therefore bind on flaws that have not yet been disclosed, in which case the disclosure estimates are a lower bound on the effect of access on discovery. 

\input{tables/tableA22_patching.tex}

\paragraph{Other robustness tests and caveats.} The results by type do not depend on the thresholds: defining high impact and many releases by the highest quartile of the full sample instead of the median gives similar estimates. Restricting the Chinese group to the 27 named firms gives a somewhat smaller but still large increase in disclosures, of about 90\%. Chinese firms differ from the Glasswing firms in three ways that the fixed effects do not remove. First, they are smaller: the 27 named firms disclose five vulnerabilities per month on average in the pre-launch window, against 29 for a Glasswing firm. Second, they disclose under Chinese rules that require reporting to the Ministry of Industry and Information Technology and restrict sharing with foreign organizations, and part of their flaws appear only in the Chinese national database, published in irregular batches. Third, they have access to domestic frontier models and adopt generative AI quickly \citep{baviskar2026tracking,li2026hiring}. The comparison therefore estimates the effect of access to Claude Mythos Preview relative to firms without it but with other AI tools, which is the relevant counterfactual for the question of what a frontier model adds.

\section{Researcher-Level Evidence}\label{sec:researchers}

The firm-level evidence shows that the Glasswing firms found and disclosed many more vulnerabilities. This section asks what happened to the researchers who used to find vulnerabilities in their code. No researcher outside the Glasswing firms received access to Claude Mythos Preview. External researchers were affected only because the vendors whose code they search acquired a frontier model and began to find flaws themselves, competing with them for the same stock of undiscovered vulnerabilities. Project Glasswing is therefore a shock to the competition that external researchers face, which differs across researchers according to whose code they worked on.

\paragraph{Design choice.} Two specifications are possible. The first follows \citet{berg2021spillover} and estimates, on all researchers, the direct effect on the researchers who work for a Glasswing firm and have direct access, and the spillovers on researchers with more or less exposure to Glasswing code. The second restricts the sample to external researchers and compares researchers with different exposure. I use the second as the main specification for three reasons. First, the direct effect is poorly measured: Glasswing firms credit their own teams for only a fraction of internal finds, and the finder of a quarter of the Glasswing firms' vulnerabilities is unknown (Table~\ref{tab:t1}), so the credited output of researchers with direct access understates what they found with the model. Second, the economically new question is the response of the external market, the researchers whose work AI can substitute for and who can choose where to search; the effect on the Glasswing firms' own output is already measured at the firm level. Third, the external sample avoids comparing very different groups, the security teams of the largest technology firms and independent researchers.

\paragraph{Sample and exposure.} The sample consists of the 7,237 external researchers, persons and teams, who were credited with at least one vulnerability published in 2024 and are not affiliated with a Glasswing firm. I follow them every month from January 2025 to September 2026; months without credits enter as zeros, so leaving the market is an outcome. Exposure is measured in 2024, a year before the panel starts. For each researcher, $s^{G}_i$ is the share of the vulnerabilities credited in 2024 (each counted $1/N$ when $N$ finders are credited) that affected the Glasswing firms, $s^{O}_i$ the share at other non-Chinese vendors and $s^{C}_i$ the share at Chinese vendors. The three shares sum to one. Measuring exposure before the panel matters: a researcher selected because of many Glasswing vulnerabilities in a given period will, on average, report fewer of them afterwards whatever happens, and a gap between the exposure window and the panel removes this mechanical decline from the comparison of pre-launch and post-launch months. The timing of exposure differs across the three types of code. Researchers who worked on Glasswing code faced a competitor with the model from April 2026, and from May 2026 in IBM's code; those who worked on other non-Chinese code faced competitors with access only after the expansion of June 2026; and those who worked on Chinese code never did, because Chinese firms cannot use Anthropic's models. On average, 14\% of an external researcher's 2024 vulnerabilities were at Glasswing firms, 83\% at other non-Chinese vendors, and 3\% at Chinese vendors; 981 researchers had more than half of their vulnerabilities at Glasswing firms and 185 at Chinese vendors.

\paragraph{Specification.} I estimate by Poisson pseudo-maximum likelihood
\begin{equation}
\mathbb{E}\left[V_{it}\mid\cdot\right] = \exp\left(\alpha_i + \delta_t + \beta_G\, s^{G}_i\times\text{Post}_t + \beta_O\, s^{O}_i\times\text{Expansion}_t\right),
\label{eq:researcher}
\end{equation}
where $V_{it}$ is the number of vulnerabilities credited to researcher $i$ in month $t$, in total or in a given type of code, $\alpha_i$ and $\delta_t$ are researcher and year-month fixed effects, $\text{Post}_t$ equals one from April 2026 (for the share of a researcher's exposure that comes from IBM's code, from May 2026) and $\text{Expansion}_t$ from June 2026. The share of Chinese code is the omitted category, so $\beta_G$ compares a researcher who worked only on Glasswing code with one who worked only on Chinese code, and $\beta_O$ does the same for other non-Chinese code after the expansion. Because exposure comes from the code of a limited number of vendors, I cluster standard errors by the researcher's main issuing organization in 2024, the level at which exposure varies most, which gives 166 clusters. Splitting a researcher's credits by the type of code shows whether a fall in Glasswing vulnerabilities is matched by more vulnerabilities elsewhere (reallocation) or not (crowding out and exit).

\paragraph{Results.} Table~\ref{tab:res} reports the estimates. External researchers exposed to Glasswing code found fewer vulnerabilities there after the launch: a researcher who had worked only on Glasswing code reported 57\% fewer Glasswing vulnerabilities than a researcher with no such exposure (column 2). For an individual researcher the effect is small in levels; for the 981 researchers with more than half of their 2024 work on Glasswing code, it amounts to about one fewer Glasswing vulnerability every two years. The researchers also reported about two and a half times as many vulnerabilities in the code of other non-Chinese vendors (column 3) and more in Chinese code (column 4), although the latter estimate rests on few researchers. Their total output rose, although only at the 10\% significance level (column 1), and their probability of being active in a month did not change (column 7). External researchers therefore did not leave the market; they moved out of the code in which the vendors' own AI-assisted search competed with them and into code in which it did not. AI-exposed and other vulnerabilities rose by similar amounts (columns 5 and 6), so the reallocation is across vendors rather than toward flaws that AI finds less easily. Researchers who had worked on other non-Chinese code did not reduce their output after the expansion of June 2026, possibly because the expansion reached only part of that code.

\input{tables/table4_researchers.tex}

\begin{figure}[!htbp]
\centering
\includegraphics[width=0.9\textwidth]{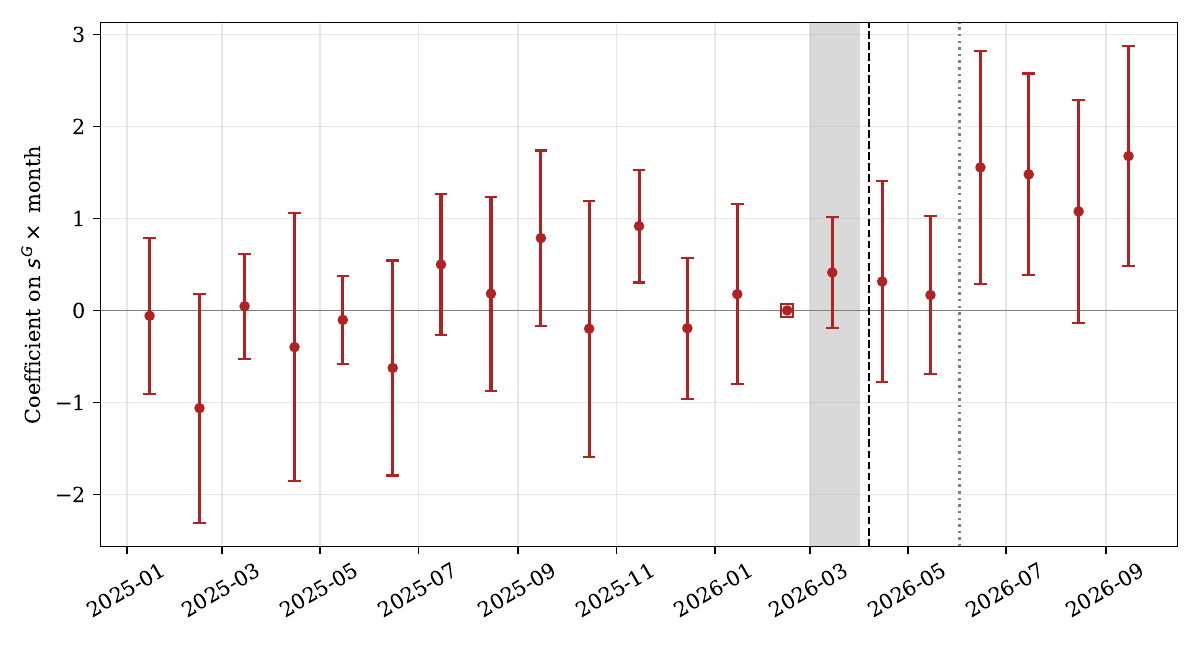}
\caption{Dynamic Difference-in-Differences: External Researchers}\label{fig:reses}
\begin{minipage}{0.92\textwidth}\footnotesize \textit{Notes:} Coefficients on $s^{G}$, the researcher's 2024 share of vulnerabilities at Glasswing firms, interacted with month indicators, from a Poisson regression of all vulnerabilities credited to the researcher in the month on these interactions, researcher and year-month fixed effects and $s^{O}\times$Expansion as a control, for the 7,237 external researchers credited in 2024, January 2025 to September 2026. The reference month is February 2026 (open square). Bars show 95\% confidence intervals based on standard errors clustered by the researcher's main issuing organization in 2024. The shaded band marks March 2026, the dashed line the launch on April 7, 2026, and the dotted line the expansion on June 2, 2026.\end{minipage}
\end{figure}

Figure~\ref{fig:reses} shows the timing. The total output of exposed researchers does not change at the launch and rises after the expansion of June 2026, when the shift toward other code accelerates. For vulnerabilities in Glasswing code, the monthly coefficients are negative after the launch, but they are also mostly positive in the first half of 2025 and drift toward zero during that year. Part of the decline in Glasswing vulnerabilities may therefore continue a pre-existing drift, the remaining mean reversion of researchers selected on their 2024 activity, and the monthly coefficients are imprecise.

\paragraph{Interpretation and caveats.} The evidence points to reallocation rather than exit. A frontier model in the hands of the vendors lowered the expected return to searching their code for external researchers, who responded by moving to code where the vendors did not have it. This is the substitution of AI for external human labor at the level of the codebase, combined with an elastic reallocation of that labor across codebases, and it suggests that the security benefits of the model spill over to the rest of the software ecosystem through researchers' reallocated effort. Three caveats apply. First, the credited vulnerabilities measure output, not effort, and the returns to research, which the repricing of bug bounties suggests fell, are not yet measured. Second, exposure varies across a limited number of vendors, so the effective number of independent shocks is small even though the number of researchers is large. 

\section{Attackers: Exploitation of Vulnerabilities}\label{sec:attack}

Attackers did not have access to Claude Mythos Preview, but they observe every vulnerability that the Glasswing firms publish. A disclosure turns a secret flaw into public information, and both disclosure and the release of a patch raise the frequency of attacks in the short run \citep{arora2006does} and bring them forward in the life of a vulnerability \citep{mitra2015information}. Many users install patches slowly \citep{murciano2024navigating}, so more disclosures could mean more exploitable vulnerabilities in the installed base. Against this, a model that searches code for latent flaws finds vulnerabilities that attackers have not found, and fixing them removes them from attackers' reach. This section asks which force dominates in the first months after the launch.

\paragraph{Measurement.} Disclosure can lead to two different events. Public exploit code is a working demonstration of how to use a flaw, typically published by a security researcher on GitHub or added to a widely used testing tool. It is not an attack, but it lowers the cost of attacking for anyone who wants to use the flaw. Exploitation in the wild is an actual attack on systems whose owners did not authorize it, observed by a security firm, a government agency, or the victim and recorded in the catalogs of exploited vulnerabilities described in Section~\ref{sec:sample}. Public exploit code therefore measures the supply of attack technology, and exploitation measures its use. The two are only loosely linked: many flaws with public exploit code are never used in attacks, and many attacks use code that is never published. Let $P_j$ be the publication date of vulnerability $j$, $E_j$ the date of the first public evidence that it was exploited in the wild, and $C_j$ the date on which exploit code for it first became public. For a window of $h$ days,
\begin{equation}
\text{Exploited}^{h}_{j} = \mathbf{1}\left[E_j \le P_j + h\right], \qquad \text{Code}^{h}_{j} = \mathbf{1}\left[C_j \le P_j + h\right],
\label{eq:window}
\end{equation}
with $h$ equal to 30, 60, and 90 days. I split exploitation into \emph{zero-day} exploitation, $E_j \le P_j$, in which attackers used the flaw before it was disclosed, and \emph{n-day} exploitation, $P_j < E_j \le P_j + h$, in which they responded to the disclosure. A vulnerability enters the sample for window $h$ only if $P_j + h$ falls before the end of the data (October 28, 2026), so every vulnerability is observed for the whole window. A fixed window removes the mechanical decline in exploitation for recent vulnerabilities that an ever-exploited indicator would produce, and it measures vulnerabilities published before and after the launch in the same way. As a third measure, I use the Exploit Prediction Scoring System (EPSS), which FIRST, the international association of computer security incident response teams, publishes every day for every published CVE: the predicted probability that the vulnerability will be exploited in the next 30 days, estimated from exploitation activity observed by security firms and from features of the vulnerability. Because it uses observed activity rather than confirmed reports, EPSS responds faster than the catalogs. $\text{EPSS}^{k}_{j}$ is the score at the end of the $k$-th month after the month of publication, for $k$ equal to 1 and 3, in percentage points.

\paragraph{Exploitation by group of firms.} Table~\ref{tab:exgroups} describes the exploitation of the vulnerabilities published between January 2024 and March 2026, followed to October 2026. Besides the Glasswing and Chinese firms, it shows the 26 large non-Chinese firms that form the second comparison group in Section~\ref{sec:firm} and all other firms, the thousands of remaining vendors in the CVE List, from large firms such as Samsung, Siemens, and Dell to the developers of website plugins. Chinese products are not less attractive to attackers: about 1.6\% of Chinese vulnerabilities were exploited by October 2026, similar to the other non-Glasswing groups, against 2.4\% at the Glasswing firms. What differs is when, and by whom, the exploitation is detected. The exploited Chinese vulnerabilities are mostly flaws in routers and cameras, such as those of TOTOLINK and Tenda, which networks of infected devices use at scale. Only a quarter of them were already exploited when published, against more than half at the Glasswing firms, and the first evidence arrives a median of 175 days after publication, against the day of publication for the Glasswing firms. Few of them appear in the catalog of the US Cybersecurity and Infrastructure Security Agency, which focuses on the networks of the US government; most of the evidence comes from VulnCheck. Within 90 days of publication, exploitation is therefore detected for only 0.6\% of Chinese vulnerabilities, about half the rate of the large non-Chinese firms. Attacks on Chinese products are observed later, not less often.

\input{tables/table_exploitation_groups.tex}

\paragraph{Comparison groups.} As in Sections~\ref{sec:firm} and~\ref{sec:researchers}, the first comparison group consists of Chinese firms, from October 2025 to the last month each window allows. The difference-in-differences design removes the level differences in Table~\ref{tab:exgroups}, but the slower detection leaves few exploited Chinese vulnerabilities within short windows: in the window used in Section~\ref{sec:firm}, eight vulnerabilities of Chinese firms were exploited within three months of publication, against 59 at the Glasswing firms. I therefore also compare the Glasswing firms with the 26 large non-Chinese firms, whose exploited vulnerabilities are detected about as quickly as those of the Glasswing firms (77\% and 76\% within 90 days), from January 2024 to May 2026, the window in which they did not yet have access; IBM is treated from May 2026. Appendix Table~\ref{tab:explcount} estimates the number of exploited vulnerabilities relative to this second group; with Chinese firms, Poisson regressions of exploited counts are not identified for some outcomes.

\paragraph{Specification.} For vulnerability $j$ attributed to firm $i$ I estimate the linear probability model
\begin{equation}
y_{ji} = \alpha_i + \delta_{t(j)} + \gamma_{c(j)} + \beta\,\text{Glasswing}_i\times\text{Post}_{t(j)} + X_j'\theta + \varepsilon_{ji},
\label{eq:cve}
\end{equation}
where $y_{ji}$ is one of the outcomes in equation~\eqref{eq:window} in percentage points, $t(j)$ is the publication month, $\gamma_{c(j)}$ are fixed effects for the weakness type and $X_j$ contains the severity score and an indicator for flaws that can be exploited over the network. A vulnerability attributed to two firms enters once for each. The probability answers whether the marginal disclosed flaw is attractive to attackers. Whether attacks on a firm's products changed is a question about the number of exploited vulnerabilities, which I estimate on the firm-month panel by Poisson pseudo-maximum likelihood with firm and year-month fixed effects \citep{silva2006log}, the natural estimator for rare counts. Standard errors are clustered by firm.

\paragraph{The probability of exploitation.} The first question is whether the vulnerabilities disclosed after access were as likely to be exploited as earlier ones. They were not. Relative to Chinese firms, the probability that a Glasswing vulnerability was exploited within a month of publication fell by about two thirds of its pre-launch rate of 1.4\%, and the probability that it was already exploited when published fell by about seven tenths (Table~\ref{tab:expl}, Panel A). Over three months the decline is smaller and not significant. The comparison with the 26 large non-Chinese firms (Panel B) uses vulnerabilities that had all been public for at least 150 days by the end of the data, so the windows differ only in how long attackers are given. It shows the same pattern: within three months, the exploitation probability fell by about two thirds of its pre-launch rate of 1.9\%, mostly because a smaller share of Glasswing vulnerabilities was already exploited when published; over one month the decline is not significant. With thirteen treated firms, clustered standard errors may overstate the precision of these estimates, so I read the decline as suggestive rather than established.

\paragraph{The number of exploited vulnerabilities.} A lower probability can reflect fewer attacks or more disclosures of flaws that attackers were not using. The number of exploited vulnerabilities distinguishes the two, and it did not change relative to the large non-Chinese firms (Appendix Table~\ref{tab:explcount}): neither the number exploited within one, two, or three months of publication nor the number of zero-day or n-day exploitations moved significantly. Before and after the launch, the Glasswing firms disclosed about 6.5 vulnerabilities a month that were exploited within three months, while their total disclosures rose by two thirds or more. The lower probability therefore reflects dilution, and there is no evidence that the vulnerabilities added by AI access were exploited, either before they were disclosed or in the three months after. The counts are imprecise, however: they rule out a doubling of attacks on the Glasswing firms but not a moderate change in either direction.

\input{tables/table7_exploitation.tex}

\paragraph{Public exploit code.} Public exploit code behaves differently from exploitation. Relative to the large non-Chinese firms, the probability that a Glasswing vulnerability had public exploit code within one or three months did not change, so the number of Glasswing vulnerabilities with exploit code rose roughly in proportion to disclosures, by about two thirds (Appendix Table~\ref{tab:explcount}). Relative to Chinese firms, exploit code rose less than disclosures. Writing exploits, much of it by researchers who publish demonstration code, keeps pace with what is disclosed, while exploitation in the wild does not. This is the pattern expected if attackers have limited capacity and choose among disclosed flaws by expected return \citep{becker1968crime}.

\paragraph{Predicted exploitation.} EPSS asks whether the additional vulnerabilities looked dangerous when they were disclosed. They did: their predicted probability of exploitation one and three months after publication did not change relative to either comparison group (Table~\ref{tab:expl}, columns 6 and 7). The additional vulnerabilities therefore looked as likely to be exploited as earlier ones, so the expected number of attacks rose with the number of disclosures, while the number of observed exploitations did not. Two caveats apply. Early scores rest largely on features of the vulnerability, such as its weakness type and whether exploit code exists, so they partly measure how dangerous a flaw looks rather than attacks on it. And EPSS changed its model in June 2026, but scoring the same vulnerabilities just before and after the change moves their mean score little and similarly for Glasswing and Chinese vulnerabilities.

\paragraph{Zero-days and pre-emption.} A natural reading of Project Glasswing is pre-emption: Glasswing firms with better tools find latent flaws before attackers do, which leaves fewer flaws for zero-day exploitation. Pre-emption predicts that the \emph{number} of zero-days in Glasswing products falls, not only their share among disclosed vulnerabilities, and the share can fall without any pre-emption. A flaw found by scanning code was, by construction, not found because it was being exploited, while many zero-days are disclosed precisely because an attack was detected. The weakness-type fixed effects and severity controls hold the type and severity of the flaws constant, but not the way in which they were found, so a large inflow of flaws found by scanning lowers the zero-day share almost mechanically. The evidence so far therefore shows that the flaws added by AI access were latent flaws that attackers were not using, which is the premise of pre-emption, but not pre-emption itself: the number of Glasswing zero-days did not change (Appendix Table~\ref{tab:explcount}). Pre-emption, if it exists, should appear with a lag, because zero-days exploited between April and September 2026 were largely developed before the launch, and the Glasswing firms have only about four zero-days a month, so even a large proportional fall is hard to detect within six months. The time from disclosure to exploitation cannot yet be estimated: only five Glasswing CVEs from the launch window were first exploited after disclosure and within 90 days.

\paragraph{Caveats.} Exploitation is rare: Glasswing firms average about 0.23 CVEs exploited within three months per firm and month, so the count regressions can rule out only large changes. The dates record when exploitation became publicly known, not when it began, and part of the detection is done by the Glasswing firms' own teams that monitor attacks. Using CISA's catalog alone, which depends less on any single vendor's reporting, gives the same pattern: the probability falls by a similar amount, and the count does not change. Windows of up to three months capture early exploitation only; because the windows are fixed, longer windows can be added as the data are updated without changing the design.

\section{Conclusion}
\label{sec:con}
Frontier AI models can now find software flaws at a scale that used to require large teams of skilled researchers. Whether such task-level gains raise security depends on what happens around the task: whether the additional output is valuable, whether downstream human work can absorb it, how producers without access respond, and how third parties use the information. In cybersecurity the last channel is adversarial, because every disclosure that helps defenders also informs attackers. Project Glasswing, which gave a frontier model to a small group of large US firms while the provider's terms barred Chinese and Russian firms from it, allows me to study each of these channels.

Relative to Chinese firms, firms with access more than doubled their disclosures within six months, and the increase was largest for high-impact flaws. The additional vulnerabilities were new discoveries rather than an old backlog, a third or more of the increase came from the firms' own teams, and vendors and Linux distributions fixed reported flaws as quickly as before. External researchers who had searched the treated firms' code reported about half as many vulnerabilities there, but they did not leave the market; they moved to code that the model had not reached. And there is no evidence that attackers exploited the additional vulnerabilities: the share exploited fell because the added flaws were latent ones that attackers were not using, while the number of exploited vulnerabilities did not change.

These results point to a general feature of AI in adversarial settings. When defenders and attackers search the same stock of flaws, the value of a better search technology depends less on the number of flaws it finds than on whether it finds them first. Access to the model let defenders remove latent flaws before attackers used them, and the researchers it displaced carried their effort to other code, so part of the benefit spread beyond the firms with access. The evidence also points to where the next constraint lies. Anthropic reports that its partners found far more flaws than they have disclosed, so fixing, not finding, may become the binding constraint, and the estimates are a lower bound on the effect of access on discovery. A fix also protects only the users who install it: because many users patch slowly, the security gain depends on how quickly fixes reach the systems in use, a stage this paper does not observe. Finally, whether defenders stay ahead depends on attackers: once comparable models reach them, the race is decided by the relative capabilities of the two sides, not by the capability of defenders alone.

The allocation of access has a geopolitical dimension. The firms excluded from the model were Chinese and Russian, and relative to them the firms with access roughly doubled the latent flaws they removed from their software. The restriction was set by a private firm rather than a government, but it works like an export control: it allocates a strategic technology along geopolitical lines. If frontier models continue to be released first to selected firms in selected countries, the security of a country's software will depend not only on its ability to build such models but also on its access to them. This raises questions that the evidence here can only begin to answer: how long the security gap persists once models diffuse and whether excluded countries respond by building their own defensive models.

\newpage
\bibliographystyle{econ}
\bibliography{literature}

\newpage
\appendix
\setcounter{table}{0}\renewcommand{\thetable}{A\arabic{table}}
\setcounter{figure}{0}\renewcommand{\thefigure}{A\arabic{figure}}

\section{Data}\label{sec:construction}

This appendix describes the technical steps behind the samples in Section~\ref{sec:data}, lists the firms in the sample and examines whether Chinese disclosure changed around the launch. 

\subsection{CVE records and attribution to firms}

\begin{itemize}
\item \emph{CVE List.} Official CVE List in CVE JSON 5 format, snapshot of October 28, 2026 (416,025 records, of which 397,187 are published). For each CVE: the issuing CVE Numbering Authority (CNA), the reservation and publication dates, the affected vendors and products reported by the issuer, the reference URLs, the credited parties with their roles, and, from CISA's Authorized Data Publisher container, CISA's exploitation and automatability assessments (SSVC) and listing in CISA's Known Exploited Vulnerabilities (KEV) catalog.
\item \emph{National Vulnerability Database (NVD).} Downloaded through the NVD API and refreshed on October 28, 2026, for all CVEs published since January 2024: affected-product configurations (CPE), weakness types (CWE) and severity scores (CVSS). Weakness types and severity fall back on the issuer's or CISA's assessment when the NVD has none.
\item \emph{Attribution.} A CVE is attributed to a firm if (i) the firm is a CPE vendor of the CVE, (ii) the firm is the issuing CNA and the affected vendors do not belong to a different firm, (iii) the firm is an affected vendor reported by the issuer, or (iv) the CVE references the firm's own security advisory pages (for example \texttt{msrc.microsoft.com}, \texttt{support.apple.com}, Chrome release notes and Android bulletins for Google, Oracle's Critical Patch Update pages). Research groups that publish other vendors' vulnerabilities (Cisco Talos, Google Project Zero, the research teams of Tenable, Rapid7, and Qualys) do not attribute a CVE.
\item \emph{Platform rule.} The NVD lists the platforms on which a vulnerable application runs as additional operating-system (\texttt{o}) or hardware (\texttt{h}) CPEs. A firm whose CPEs on a CVE are all of these two types is not attributed the CVE when the CVE also has an application (\texttt{a}) CPE of another vendor.
\item \emph{Timing of NVD data.} The CVE-List sources matter because NVD enrichment lags publication: in the NVD download of September 10, 2026, only 71\% of the Glasswing CVEs published after the June expansion had a CPE configuration, against 93\% after the refresh of October 28, 2026.
\end{itemize}

\subsection{Exploitation and exploit code}

\begin{itemize}
\item \emph{First exploitation.} Earliest of CISA KEV's date added, VulnCheck KEV's date added, and VulnCheck KEV's date of first reported exploitation evidence.
\item \emph{First public exploit code.} Earliest of the creation date of the first proof-of-concept repository on GitHub (PoC-in-GitHub index), the date on which a module of the Metasploit attack framework or a template of the Nuclei vulnerability scanner for the CVE was added (from the git history of each project) and the date of the first exploit in VulnCheck's exploit index.
\item \emph{Zero-day and n-day.} A CVE is exploited as a zero-day if the first exploitation date is on or before its publication date, and as an n-day if it is after.
\end{itemize}

\subsection{Sources of finder information}\label{sec:sources_app}

Table~\ref{tab:sources} lists each source with the number of sample CVEs it covers and the share of those CVEs for which it names at least one finder.
\begin{enumerate}
\item \emph{The CVE List.} The credits field of each CVE record, with the role of each credited party (finder, reporter, analyst, tool, coordinator, remediation developer).
\item \emph{Vendor acknowledgments.} Microsoft (Security Update Guide), Google (Chrome release notes and Android bulletins), Apple, Cisco, NVIDIA, Adobe, Oracle, IBM, Intel (machine-readable advisories on GitHub), Qualcomm, Fortinet, Dell, Huawei, Samsung Mobile, Red Hat, Siemens, and JPCERT/CC's Japan Vulnerability Notes. Some vendors also state when a flaw was found internally (``found during internal security testing,'' ``found internally by Intel employees'').
\item \emph{Disclosure and bounty programs.} Trend Micro's Zero Day Initiative advisories and HackerOne's disclosed reports.
\item \emph{Open-source advisories.} The GitHub Advisory Database, the credits of GitHub repository security advisories (through GitHub's API) and OSV.dev.
\item \emph{Fixing commits.} ``Reported-by'' tags in the commits that fix a flaw, from the mainline Linux kernel history (matched to the fixing commit of every kernel CVE) and from the fixing commits on GitHub referenced by other CVEs. The email domain identifies the reporter's organization; automated reporters such as Google's syzbot fuzzer are recorded as tools.
\item \emph{References and discovery statements.} When no other source names a finder, a reference to a researcher's own write-up (a GitHub repository or gist, a Medium, GitHub Pages, Blogspot, or Substack post) identifies the researcher by account name, and a reference to a research team's advisory page (Cisco Talos, the Google Project Zero tracker, GitHub Security Lab) identifies the team. Some issuers fill the discovery field of the CVE record, which states whether the flaw was found internally or externally without naming the finder.
\end{enumerate}
Coverage differs in kind across sources. Vendor acknowledgment pages cover only the vendor's own CVEs and, for most vendors, name a finder for almost every externally reported flaw; a low share for a vendor (Oracle, IBM, Dell, Huawei, Red Hat, Siemens) means that most of its CVEs were found internally or that its advisories credit researchers only for a whole patch release. For Dell, Huawei, Red Hat, and Siemens the pages were collected only for CVEs without a finder from any other source. The open-source databases mirror most open-source CVEs but add names for few of them, while the credits of GitHub repository advisories name a finder for 70.5\% of the CVEs GitHub issues. Fixing commits add reporters for a quarter of Linux kernel CVEs; the remaining kernel fixes carry no Reported-by tag, which in kernel practice means that the flaw was found by the developer who fixed it or by a reviewer.

\input{tables/tableA5_hunter_sources.tex}

\subsection{From credits to researcher identifiers}

Each source yields credit strings such as ``Hongli Han (@hexb1n) and Guang Gong (@oldfresher) of 360 Alpha Lab working with 360 BugCloud'' or ``Cisco would like to thank X of Y for reporting this vulnerability.'' I convert them into researcher identifiers in six steps.
\begin{enumerate}
\item \emph{Roles.} Finders, reporters, analysts, and tools are kept; coordinators, remediation developers and reviewers, sponsors, and distributions that only repackage fixes are dropped.
\item \emph{Parsing.} Boilerplate (``would like to thank'', ``for reporting this issue'') is removed; a credit is split into entities at ``and'', ``\&'', commas, and semicolons; each entity's name is separated from its handle (in parentheses or after ``@'') and its affiliation (after ``of'', ``from'', ``at'', a dash, or a trailing organization name). An affiliation stated once for a list of names applies to all of them. Intermediaries (``working with Trend Micro Zero Day Initiative'', ``via HackerOne'') and AI tools named as instruments (``using Claude'') are recorded separately and are not the finder's employer.
\item \emph{Entity type.} Person, team or organization, AI system (``Claude'', ``Google Big Sleep''), automated tool (syzbot), anonymous researcher, or, for Microsoft's hashed reporter names, pseudonymous identifier. A statement of internal discovery without names becomes the internal team of the affected vendor (``Cisco internal team'').
\item \emph{Standardized key and identifier.} The key is the entity type and the standardized name (accents removed, case folded, punctuation other than ``@'', ``.'', ``-'', and ``\_'' removed, spaces collapsed), so that ``Ra\'ul S\'anchez'' and ``raul sanchez'' coincide. The identifier is the first ten hexadecimal characters of the SHA-1 hash of the key, so that the same name receives the same identifier whenever the data are rebuilt. All anonymous credits share one identifier.
\item \emph{Same person, different names.} For each identifier I collect the handles, the GitHub accounts, credited names that are themselves account names (a single word of at least four characters), and the GitHub accounts matched to the 500 most prolific researchers by name, handle, and profile, matches of high or medium confidence). Identifiers that share one of these are merged (union-find), and the merged person keeps the identifier with the most credits. An account shared by more than five identifiers, an organization name, or a web domain is not used, because colleagues share them. This step merges 814 identifiers into 712 persons.
\item \emph{Organizations.} A researcher's affiliation is the organization it is most often credited with (ties go to the most recent). Organization names are matched to parent firms by name patterns: Google Project Zero, Mandiant, and Big Sleep map to Google; Talos and Splunk to Cisco; MSRC, MORSE, and GitHub Security Lab to Microsoft; Unit 42 to Palo Alto Networks. A researcher is affiliated with a treated firm if its affiliation is a launch partner.
\end{enumerate}
The database contains 275,385 links between researchers, CVEs, and sources across all years.

\subsection{Finder types and inferred internal discovery}

The four finder types in Section~\ref{sec:measures} are assigned in order: internal if any finder works for the affected vendor (including vendor statements and the CVE record's discovery field set to internal), otherwise another organization or project if any finder is credited with an organization, as a team, an AI system, or a tool, otherwise an individual researcher if any finder is acknowledged, otherwise unknown. Table~\ref{tab:inferred} adds two inferred categories to the internal finds: CVEs listed on the vendor's own acknowledgment page without any credit when the vendor issued the CVE (7,186 CVEs, mostly of Red Hat, Siemens, Dell, and Huawei), and Linux kernel fixes without a Reported-by tag (11,742 CVEs).

\input{tables/tableA7_inferred_internal.tex}

\subsection{AI exposure}

Table~\ref{tab:adcsum} lists the weakness types (CWE) behind the AI-exposure measure. Memory-safety weaknesses are CWE-119 to CWE-127, 129 to 131, 134, 170, 190, 191, 193, 415, 416, 457, 466, 476, 562, 587, 590, 680, 690, 761 to 763, 786 to 788, 805, 806, 822 to 825, 908, and 909. Machine-verifiable weaknesses add injection (CWE-74 to 98, 113, 116, 117, 470, 564, 643, 652, 917, 943, 1336), path traversal (CWE-22 to 40, 73), unsafe deserialization (CWE-502), XML external entities (CWE-611) and server-side request forgery (CWE-918). A CVE with several weakness types is AI-exposed if any of them is a memory-safety weakness.

\input{tables/tableA2_classification.tex}

\subsection{Russian firms}

Russian firms form a further comparison group in Table~\ref{tab:firm}, columns (7) and (8). Anthropic's terms bar companies controlled from Russia, as from China. A CVE belongs to a Russian firm if the firm is named as the affected vendor or as the vendor in the product configuration of the National Vulnerability Database, if the firm issued the identifier for its own product, or if the description names the firm's product. Kaspersky also issues identifiers for flaws in other vendors' industrial software, so Kaspersky CVEs are attributed by vendor only. Firms founded in Russia but headquartered elsewhere are not Russian firms (ONLYOFFICE, MikroTik, JetBrains, Acronis, ClickHouse, Telegram, and ABBYY). Between January 2024 and October 2026 ten Russian firms have at least one CVE: Yandex (12 CVEs), AxxonSoft (8), 1C-Bitrix (7), Kaspersky (4), Positive Technologies (2), VK (Tarantool), ROSA, New Cloud Technologies (MyOffice), ISPsystem, and R7 (one each); nine of them have a vulnerability in the analysis sample. Several Russian vendors that publish their vulnerabilities in the national database of the Federal Service for Technical and Export Control rather than as CVEs, such as Red Soft, Astra Linux, BaseALT, Postgres Professional, and UserGate, have no CVE in the period.

\FloatBarrier

\subsection{Firms in the sample}\label{sec:firmlist}

Table~\ref{tab:firms} lists the firms in the firm-level and attacker samples with their average number of vulnerabilities per month before the launch, in the first two months after it and after the expansion.

\input{tables/tableA1_firms.tex}
\FloatBarrier

\subsection{Chinese disclosure behavior}\label{sec:chinese}

The main comparison group consists of Chinese firms, whose disclosure is regulated by the state (Section~\ref{sec:disclosure}). If Chinese authorities had tightened disclosure after the launch of Project Glasswing, for instance by holding back vulnerabilities in Chinese products, the disclosures of Chinese firms would have fallen for reasons unrelated to access to the model, and the estimates would overstate the effect. This subsection examines three kinds of evidence: the channels through which Chinese vulnerabilities are disclosed, the behavior of independent researchers who study Chinese products, and the timing of disclosure. Table~\ref{tab:chinese} reports monthly means for 2024, the six months before the launch and the six months after it.

\begin{figure}[!htbp]
\centering
\includegraphics[width=\textwidth]{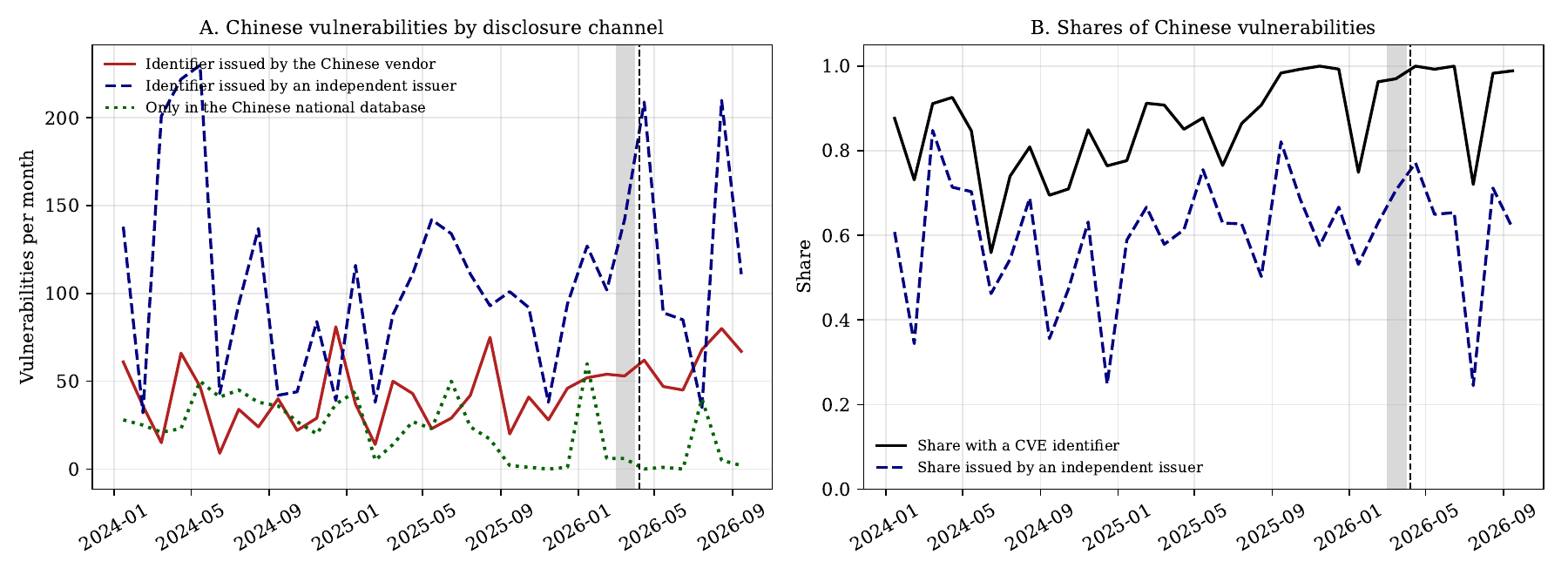}
\caption{Chinese Vulnerabilities by Disclosure Channel}\label{fig:chinesechannels}
\begin{minipage}{0.95\textwidth}\footnotesize \textit{Notes:} Panel A: vulnerabilities of Chinese firms published each month, January 2024 to September 2026, split into those whose CVE identifier was issued by the Chinese vendor itself, those whose identifier was issued by an independent issuer (mainly VulDB and MITRE), and entries published only in the China National Vulnerability Database. Panel B: shares of all Chinese vulnerabilities in the month with a CVE identifier and with an identifier from an independent issuer. The shaded band marks March 2026 and the dashed line the launch on April 7, 2026.\end{minipage}
\end{figure}

\paragraph{Disclosure channels.} A vulnerability in a Chinese product can become public in three ways: the Chinese vendor issues a CVE identifier for it, an independent issuer such as VulDB or MITRE issues one, usually on a report by a researcher, or the vulnerability appears only in the China National Vulnerability Database. The Chinese rules bind most directly on the vendors and on the national database. An independent issuer does not need the vendor's cooperation, and researchers outside China are not bound by the Chinese rules, so a tighter policy would have reduced the vulnerabilities that Chinese vendors publish themselves and raised the share published through independent issuers. Figure~\ref{fig:chinesechannels} shows neither. Vulnerabilities whose identifier the Chinese vendor issued rose from 46 to 62 a month after the launch, those with an identifier from an independent issuer from 99 to 123, and the share published through independent issuers was 0.63 before the launch and 0.61 after it. A shift from the CVE List toward domestic disclosure would also have raised the share of entries published only in the national database; instead, the share of Chinese vulnerabilities with a CVE identifier was 0.94 before the launch and 0.95 after it, and the share of high- or critical-severity vulnerabilities 0.66 and 0.64. Two features of the series predate the launch or are unrelated to China. Entries published only in the national database fell from about 33 a month in 2024 to almost none from September 2025, seven months before the launch, so this channel cannot reveal a further restriction after April 2026. Vulnerabilities with an identifier from an independent issuer dipped from May to July 2026, when VulDB issued fewer identifiers for Chinese products (85, 34, and 18 against 172 in April), and recovered in August.

\begin{figure}[!htbp]
\centering
\includegraphics[width=\textwidth]{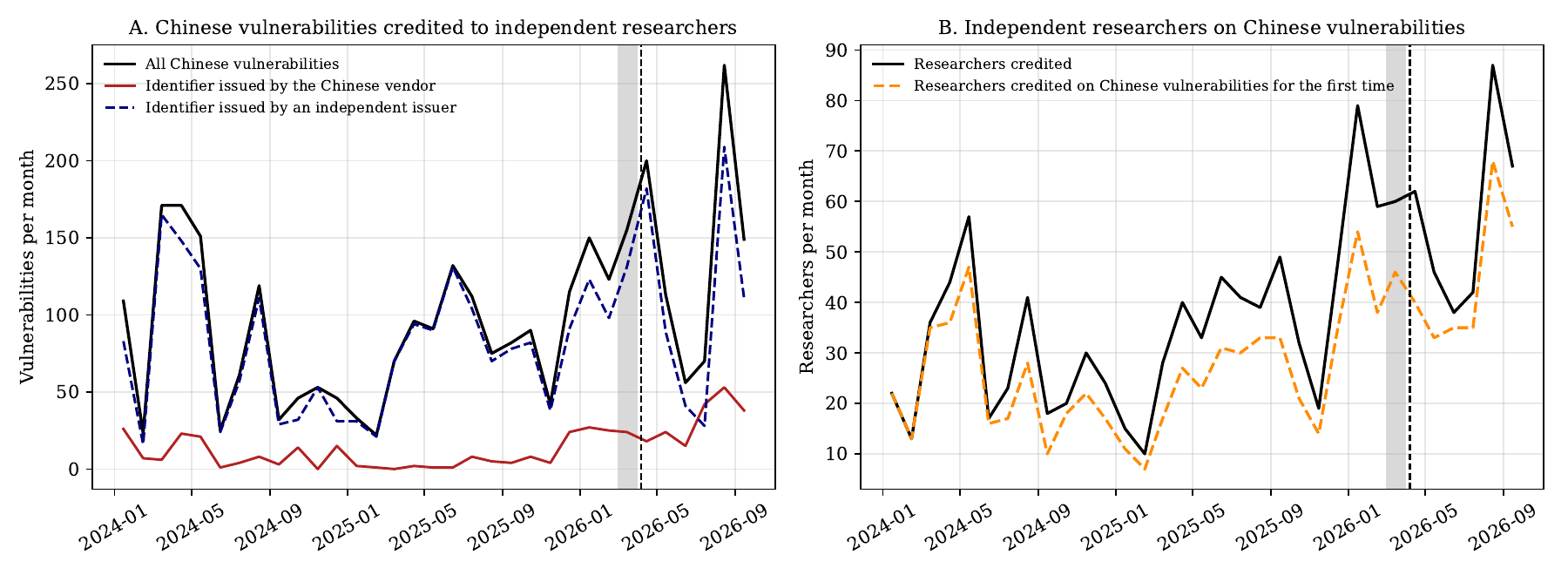}
\caption{Independent Researchers on Chinese Vulnerabilities}\label{fig:chineseres}
\begin{minipage}{0.95\textwidth}\footnotesize \textit{Notes:} Panel A: vulnerabilities of Chinese firms credited to at least one independent researcher each month, January 2024 to September 2026, in total and by issuer. Independent researchers are persons, teams, and pseudonymous identifiers credited on the vulnerability record, other than the vendor's own staff and researchers affiliated with a Glasswing firm. Panel B: independent researchers credited on Chinese vulnerabilities in the month, and those whose first credit on a Chinese vulnerability since January 2024 falls in the month; the second series is mechanically high in early 2024. The shaded band marks March 2026 and the dashed line the launch on April 7, 2026.\end{minipage}
\end{figure}

\paragraph{Independent researchers.} A tighter disclosure regime would also have lowered the return to studying Chinese products for independent researchers, whose credit and bounties depend on publication. The opposite happened. Chinese vulnerabilities credited to an independent researcher, a person, team, or pseudonymous identifier other than the vendor's own staff and researchers affiliated with a Glasswing firm, rose from 113 to 142 a month after the launch. The number of independent researchers credited on Chinese vulnerabilities rose from 50 to 57 a month, and the number credited on a Chinese vulnerability for the first time since January 2024 from 35 to 44 (Figure~\ref{fig:chineseres}). The rise includes the vulnerabilities that Chinese vendors publish themselves, the channel the rules cover most directly: these credited an independent researcher 32 times a month after the launch against 19 times before it. These series have risen since early 2025 and are volatile, so I read them as evidence against a restriction rather than as an effect of the launch; the reallocation of exposed researchers toward other code is estimated in Section~\ref{sec:researchers}.

\paragraph{Timing.} A restriction could also delay disclosure rather than prevent it. The median Chinese vulnerability with a CVE identifier was published 1.6 days after its identifier was reserved in the six months before the launch and 8.1 days in the six months after it, against 7.9 days in 2024. In the national database, the median entry was published 58 days after submission before the launch and 53 days after it. These timing measures are weak evidence. Most Chinese CVE identifiers are reserved and published almost at once, and most entries in the national database (88\% before and 93\% after the launch) concern vulnerabilities already published in the CVE List, so the delay in the database mostly measures how quickly it records vulnerabilities that are already public. The means of both measures are dominated by occasional batches of identifiers reserved or submitted years earlier, such as the publication in December 2024 of identifiers that Huawei had reserved in 2019 and 2020, and the entry of router vulnerabilities from 2022 to 2024 into the national database in late 2025, and are not informative.

Taken together, the evidence does not point to a change in Chinese disclosure around the launch. It cannot rule out that Chinese authorities withheld vulnerabilities that never reached any public database, which no public data can reveal.

\input{tables/tableA_chinese_disclosure.tex}

\FloatBarrier

\clearpage
\section{Additional Tables and Figures}
\input{tables/tableA18_firm_loo.tex}
\input{tables/tableR_inference.tex}
\input{tables/tableA_opus.tex}
\input{tables/tableA11_firm_gap.tex}
\input{tables/tableA14_firm_finder.tex}
\input{tables/table10_exploited_counts.tex}

\end{document}

%% file: tables/table1_sample.tex
\begin{table}[!htbp]\centering
\caption{Descriptive Statistics}\label{tab:t1}
\small
\resizebox{\textwidth}{!}{%
\begin{tabular}{lcccccc}
\toprule
\multicolumn{7}{l}{\textit{Panel A: vulnerabilities by group of firms}} \\[2pt]
 & \multicolumn{3}{c}{Full sample} & \multicolumn{3}{c}{Pre-treatment (Jan 2024--Mar 2026)} \\
\cmidrule(lr){2-4}\cmidrule(lr){5-7}
 & & & & Glasswing & Other non- & Chinese \\
 & Obs. & Mean & SD & firms & Chinese firms & firms \\
\midrule
Observations (pre-treatment) & & & & 9,246 & 90,077 & 4,577 \\
Glasswing firms & 177,104 & 0.121 & 0.326 & & & \\
Other non-Chinese firms & 177,104 & 0.845 & 0.361 & & & \\
Chinese firms & 177,104 & 0.034 & 0.181 & & & \\
Exploited within 2 months & 145,493 & 0.009 & 0.094 & 0.018 & 0.009 & 0.005 \\
Exploited within 3 months & 133,500 & 0.010 & 0.098 & 0.018 & 0.010 & 0.006 \\
AI-exposed (memory-safety vulnerability) & 165,832 & 0.186 & 0.389 & 0.330 & 0.155 & 0.411 \\
Found by the vendor's internal team & 177,104 & 0.039 & 0.195 & 0.133 & 0.015 & 0.002 \\
Found by another organization or project & 177,104 & 0.165 & 0.371 & 0.272 & 0.168 & 0.127 \\
Found by an individual researcher & 177,104 & 0.527 & 0.499 & 0.344 & 0.561 & 0.588 \\
Finder unknown & 177,104 & 0.269 & 0.443 & 0.251 & 0.256 & 0.284 \\
\midrule
\multicolumn{7}{l}{\textit{Panel B: Researchers $\times$ month}} \\[2pt]
 & \multicolumn{3}{c}{Full sample} & \multicolumn{3}{c}{Pre-treatment (Jan 2024--Mar 2026)} \\
\cmidrule(lr){2-4}\cmidrule(lr){5-7}
 & & & & Treated & Non-treated & \\
 & Obs. & Mean & SD & researchers & researchers & \\
\midrule
Observations (pre-treatment) & & & & 75,141 & 428,409 & \\
Vulnerabilities (1/$N$) & 634,100 & 0.149 & 2.125 & 0.143 & 0.146 & \\
\quad at Glasswing firms & 634,100 & 0.018 & 1.363 & 0.082 & 0.000 & \\
\quad at other firms & 634,100 & 0.131 & 1.628 & 0.061 & 0.146 & \\
\quad AI-exposed & 634,100 & 0.024 & 0.787 & 0.048 & 0.016 & \\
Active (at least one vulnerability) & 634,100 & 0.061 & 0.240 & 0.090 & 0.060 & \\
Affiliated with a Glasswing firm & 634,100 & 0.033 & 0.179 & 0.147 & 0.013 & \\
Team or organization & 634,100 & 0.037 & 0.189 & 0.052 & 0.034 & \\
Unaffiliated individual & 634,100 & 0.705 & 0.456 & 0.410 & 0.757 & \\
\bottomrule
\end{tabular}}
\begin{minipage}{0.97\textwidth}\footnotesize \vspace{4pt} \textit{Notes:} Panel A covers all vulnerabilities published between January 2024 and October 2026, including the entries of Chinese firms in the China National Vulnerability Database without a CVE identifier. Exploitation is defined for vulnerabilities with a CVE identifier that were public for the full window. Panel B covers the persons and teams credited on at least one vulnerability between January 2024 and March 2026, observed monthly; vulnerabilities are counted $1/N$ when $N$ finders are credited. Treated researchers have an above-median share of pre-launch vulnerabilities at Glasswing firms. Variables are defined in Section~\ref{sec:measures}.\end{minipage}
\end{table}

%% file: tables/table2_identification.tex
\begin{table}[!htbp]\centering
\caption{Firm-Level Evidence: The Number of Vulnerabilities}\label{tab:firm}
\small
\resizebox{\textwidth}{!}{%
\begin{tabular}{lcccccccc}
\toprule
 & \multicolumn{5}{c}{Comparison group: Chinese firms} & \multicolumn{3}{c}{Alternative comparison groups} \\
\cmidrule(lr){2-6}\cmidrule(lr){7-9}
 & (1) & (2) & (3) & (4) & (5) & (6) & (7) & (8) \\
 & \shortstack{Baseline} & \shortstack{Poisson} & \shortstack{Excluding\\ Anthropic} & \shortstack{Glasswing\\ trend} & \shortstack{Placebo\\ April 2025} & \shortstack{Non-Chinese\\ firms} & \shortstack{Chinese and\\ Russian firms} & \shortstack{All\\ firms} \\
\midrule
Glasswing $\times$ Post & 0.784*** & 1.136*** & 0.862*** & 0.687*** & -0.034 & 0.339* & 0.785*** & 0.778*** \\
 & (0.206) & (0.294) & (0.206) & (0.208) & (0.102) & (0.199) & (0.206) & (0.206) \\[2pt]
Glasswing $\times$ trend (months) &  &  &  & 0.0184*** &  &  &  &  \\
 &  &  &  & (0.0069) &  &  &  &  \\[2pt]
Non-Chinese $\times$ Expansion &  &  &  &  &  &  &  & 0.291 \\
 &  &  &  &  &  &  &  & (0.202) \\[2pt]
Ibragimov-M\"uller $p$-value & 0.002 & 0.007 & 0.001 & 0.007 & 0.739 & 0.031 & 0.002 & 0.002 \\
\midrule
Dependent variable & $\log(1+N)$ & $N$ & $\log(1+N)$ & $\log(1+N)$ & $\log(1+N)$ & $\log(1+N)$ & $\log(1+N)$ & $\log(1+N)$ \\
Estimator & OLS & Poisson & OLS & OLS & OLS & OLS & OLS & OLS \\
Window start & Oct 2025 & Oct 2025 & Oct 2025 & Jan 2024 & Oct 2024 & Oct 2025 & Oct 2025 & Oct 2025 \\
Window end & Sep 2026 & Sep 2026 & Sep 2026 & Sep 2026 & Sep 2025 & May 2026 & Sep 2026 & Sep 2026 \\
Firm FE & Yes & Yes & Yes & Yes & Yes & Yes & Yes & Yes \\
Year-month FE & Yes & Yes & Yes & Yes & Yes & Yes & Yes & Yes \\
Firm-months & 3,360 & 1,188 & 3,348 & 9,240 & 3,360 & 312 & 3,468 & 3,780 \\
Clusters (firms) & 280 & 99 & 279 & 280 & 280 & 39 & 289 & 315 \\
\bottomrule
\end{tabular}}
\begin{minipage}{0.97\textwidth}\footnotesize \vspace{4pt} \textit{Notes:} Estimates of equation~\eqref{eq:firm}. The dependent variable is $\log(1+N)$, where $N$ is the number of the firm's vulnerabilities published in the month; column (2) uses $N$. Columns (1) to (5) compare the 13 Glasswing firms with 267 Chinese firms, column (6) with 26 large non-Chinese firms, and column (7) with Chinese and Russian firms; column (8) uses all groups. Column (3) excludes Anthropic, column (4) adds a linear trend for the Glasswing firms, and column (5) dates a placebo launch one year before access. Non-Chinese $\times$ Expansion equals one for the 26 non-Chinese firms from June 2026. The Ibragimov-M\"uller $p$-value tests the mean of the firm-by-firm estimates with a $t$-test with $q-1$ degrees of freedom, where $q$ is the number of Glasswing firms with an estimate. Standard errors clustered by firm in parentheses. *, **, and *** denote significance at the 10, 5, and 1 percent levels.\end{minipage}
\end{table}

%% file: tables/table3_firm_types.tex
\begin{table}[!htbp]\centering
\caption{Firm-Level Evidence: Vulnerabilities by Type}\label{tab:firmtype}
\small
\resizebox{\textwidth}{!}{%
\begin{tabular}{lcccccccc}
\toprule
 & (1) & (2) & (3) & (4) & (5) & (6) & (7) & (8) \\
 & \shortstack{All} & \shortstack{AI-exposed} & \shortstack{Not\\ AI-exposed} & \shortstack{Complex weakness} & \shortstack{Machine-\\verifiable} & \shortstack{High impact} & \shortstack{Remote,\\ no credentials} & \shortstack{Many releases} \\
\midrule
Glasswing $\times$ Post & 0.784*** & 0.642*** & 0.768*** & 0.720*** & 0.825*** & 0.919*** & 0.747*** & 0.580*** \\
 & (0.206) & (0.224) & (0.226) & (0.226) & (0.195) & (0.209) & (0.187) & (0.182) \\[2pt]
\midrule
Pre-launch mean of $N$, Glasswing firms & 29.165 & 8.405 & 20.038 & 15.506 & 12.937 & 14.468 & 4.949 & 19.076 \\
Pre-launch mean of $N$, Chinese firms & 0.589 & 0.267 & 0.321 & 0.147 & 0.441 & 0.327 & 0.124 & 0.089 \\
Firm FE & Yes & Yes & Yes & Yes & Yes & Yes & Yes & Yes \\
Year-month FE & Yes & Yes & Yes & Yes & Yes & Yes & Yes & Yes \\
Firm-months & 3,360 & 3,360 & 3,360 & 3,360 & 3,360 & 3,360 & 3,360 & 3,360 \\
Clusters (firms) & 280 & 280 & 280 & 280 & 280 & 280 & 280 & 280 \\
\bottomrule
\end{tabular}}
\begin{minipage}{0.97\textwidth}\footnotesize \vspace{4pt} \textit{Notes:} Estimates of equation~\eqref{eq:firm} for the 13 Glasswing firms and 267 Chinese firms, October 2025 to September 2026. The dependent variable is $\log(1+N)$, where $N$ is the number of the firm's vulnerabilities of the type in the column published in the month. High impact and many releases are defined relative to the median of all vulnerabilities published between January 2024 and October 2026. Standard errors clustered by firm in parentheses. *, **, and *** denote significance at the 10, 5, and 1 percent levels.\end{minipage}
\end{table}

%% file: tables/tableA22_patching.tex
\begin{table}[!htbp]\centering
\caption{Remediation Speed: Was There a Human Bottleneck?}\label{tab:patching}
\small
\resizebox{\textwidth}{!}{%
\begin{tabular}{lccccc}
\toprule
 & \multicolumn{2}{c}{\shortstack{A. Vendor fix of flaws reported\\ by the Zero Day Initiative}} & \multicolumn{3}{c}{\shortstack{B. Fix by Linux distributions\\ after publication}} \\
\cmidrule(lr){2-3}\cmidrule(lr){4-6}
 & (1) & (2) & (3) & (4) & (5) \\
 & \shortstack{$\log(1+$days$)$\\ report to fix} & \shortstack{Fixed within\\ 90 days of\\ report (\%)} & \shortstack{Fixed within\\ 30 days (\%)} & \shortstack{Fixed within\\ 90 days (\%)} & \shortstack{$\log(1+$days$)$\\ if fixed} \\
\midrule
Glasswing $\times$ Post & -0.329 & 13.394 & -3.123 & -4.495 & 0.340 \\
 & (0.232) & (21.348) & (9.155) & (3.510) & (0.320) \\[2pt]
\midrule
Pre-access mean, Glasswing vulnerabilities & 4.702 & 28.070 & 78.857 & 83.429 & 1.267 \\
Pre-access mean, other vulnerabilities & 4.589 & 35.106 & 61.004 & 74.884 & 2.335 \\
Vendor or issuer FE & Yes & Yes & Yes & Yes & Yes \\
Month FE & Report & Report & Publication & Publication & Publication \\
Vulnerabilities & 658 & 658 & 14,855 & 14,855 & 12,444 \\
Vendors or issuers (clusters) & 116 & 116 & 119 & 119 & 99 \\
\bottomrule
\end{tabular}}
\begin{minipage}{0.97\textwidth}\footnotesize \vspace{4pt} \textit{Notes:} Panel A: flaws reported to vendors by the Zero Day Initiative between October 2025 and June 2026, with vendor and report-month fixed effects. Panel B: vulnerabilities published between October 2025 and July 2026 that Debian, Ubuntu, Red Hat or SUSE track, with issuer and publication-month fixed effects; column (5) uses the vulnerabilities fixed by the end of the data. The comparison group is all other non-Chinese vulnerabilities. Standard errors clustered by issuer in parentheses.\end{minipage}
\end{table}

%% file: tables/table4_researchers.tex
\begin{table}[!htbp]\centering
\caption{Researcher-Level Evidence: External Researchers}\label{tab:res}
\small
\resizebox{\textwidth}{!}{%
\begin{tabular}{lccccccc}
\toprule
 & \multicolumn{6}{c}{Credited vulnerabilities ($1/N$) in the month} & Extensive margin \\
\cmidrule(lr){2-7}\cmidrule(lr){8-8}
 & (1) & (2) & (3) & (4) & (5) & (6) & (7) \\
 & \shortstack{All} & \shortstack{Glasswing\\ code} & \shortstack{Other non-Chinese\\ code} & \shortstack{Chinese\\ code} & \shortstack{AI-exposed} & \shortstack{Not\\ AI-exposed} & \shortstack{Not active} \\
\midrule
$s^G \times$ Post & 0.489* & -0.832** & 0.891** & 2.266** & 0.463* & 0.289 & 0.003 \\
 & (0.273) & (0.387) & (0.398) & (1.068) & (0.249) & (0.270) & (0.008) \\[2pt]
$s^O \times$ Expansion & 0.084 & 0.204 & 0.141 & 3.559** & -0.083 & 0.196 & 0.000 \\
 & (0.289) & (0.384) & (0.338) & (1.695) & (0.368) & (0.319) & (0.007) \\[2pt]
\midrule
Pre-launch mean, researchers with $s^G>0.5$ & 0.098 & 0.078 & 0.020 & 0.000 & 0.030 & 0.067 & 0.934 \\
Pre-launch mean, all external researchers & 0.160 & 0.013 & 0.145 & 0.002 & 0.020 & 0.141 & 0.953 \\
Estimator & Poisson & Poisson & Poisson & Poisson & Poisson & Poisson & OLS \\
Researcher FE & Yes & Yes & Yes & Yes & Yes & Yes & Yes \\
Year-month FE & Yes & Yes & Yes & Yes & Yes & Yes & Yes \\
Researcher-months & 46,116 & 11,109 & 39,270 & 1,660 & 13,188 & 42,210 & 151,977 \\
Clusters & 166 & 70 & 160 & 31 & 88 & 164 & 224 \\
\bottomrule
\end{tabular}}
\begin{minipage}{0.97\textwidth}\footnotesize \vspace{4pt} \textit{Notes:} Estimates of equation~\eqref{eq:researcher} for the 7,237 external researchers credited in 2024, January 2025 to September 2026. The dependent variable is the number of vulnerabilities credited to the researcher in the month, each counted $1/N$ when $N$ finders are credited. Columns (1) to (6) are estimated by Poisson pseudo-maximum likelihood; column (7) is a linear probability model. Standard errors clustered by the researcher's main issuer in 2024 in parentheses. *, **, and *** denote significance at the 10, 5, and 1 percent levels.\end{minipage}
\end{table}

%% file: tables/table_exploitation_groups.tex
\begin{table}[!htbp]\centering
\caption{Exploitation by Group of Firms}\label{tab:exgroups}
\small
\resizebox{\textwidth}{!}{%
\begin{tabular}{lcccc}
\toprule
 & Glasswing & Large non-Chinese & Chinese & All other \\
 & firms & firms (26) & firms & firms \\
\midrule
Vulnerabilities & 9,246 & 5,287 & 3,906 & 84,790 \\
Exploited by October 28, 2026 (\%) & 2.39 & 1.42 & 1.61 & 1.49 \\
Exploited within 90 days of publication (\%) & 1.83 & 1.10 & 0.61 & 0.98 \\
Public exploit code by October 28, 2026 (\%) & 4.61 & 2.53 & 2.25 & 4.75 \\
\midrule
Exploited vulnerabilities & 221 & 75 & 63 & 1,265 \\
\quad already exploited when published (\%) & 57 & 44 & 25 & 41 \\
\quad first evidence within 90 days of publication (\%) & 76 & 77 & 38 & 66 \\
\quad median days from publication to first evidence & 0 & 6 & 175 & 17 \\
\quad listed by CISA (\%) & 72 & 52 & 6 & 16 \\
\bottomrule
\end{tabular}}
\begin{minipage}{0.95\textwidth}\footnotesize \vspace{4pt} \textit{Notes:} Vulnerabilities with a CVE identifier published between January 2024 and March 2026, followed to October 28, 2026. The lower panel describes the exploited vulnerabilities of each group. Large non-Chinese firms are the 26 comparison firms of Appendix Table~\ref{tab:firms}; all other firms are the remaining vendors.\end{minipage}
\end{table}

%% file: tables/table7_exploitation.tex
\begin{table}[!htbp]\centering
\caption{Exploitation, Public Exploit Code and Predicted Exploitation}\label{tab:expl}
\small
\resizebox{\textwidth}{!}{%
\begin{tabular}{lccccccc}
\toprule
 & (1) & (2) & (3) & (4) & (5) & (6) & (7) \\
 & \shortstack{Exploited\\ 1 month} & \shortstack{Exploited\\ 3 months} & \shortstack{Zero-day} & \shortstack{Exploit code\\ 1 month} & \shortstack{Exploit code\\ 3 months} & \shortstack{EPSS\\ 1 month} & \shortstack{EPSS\\ 3 months} \\
\midrule
\multicolumn{8}{l}{\textit{Panel A: Chinese firms as comparison group}} \\[2pt]
Glasswing $\times$ Post & -0.921** & -0.783 & -0.848** & -1.527*** & -0.711 & -0.100 & 0.136 \\
 & (0.435) & (0.529) & (0.361) & (0.527) & (0.589) & (0.132) & (0.161) \\[2pt]
\midrule
Pre-launch mean, Glasswing firms & 1.419 & 1.729 & 1.197 & 2.838 & 3.636 & 0.160 & 0.298 \\
Pre-launch mean, Chinese firms & 0.575 & 0.806 & 0.575 & 1.381 & 1.611 & 0.221 & 0.476 \\
Last publication month & 2026-08 & 2026-06 & 2026-06 & 2026-08 & 2026-06 & 2026-09 & 2026-07 \\
Vulnerability-firm pairs & 10,298 & 6,620 & 6,620 & 10,298 & 6,620 & 13,177 & 8,333 \\
Clusters (firms) & 62 & 59 & 59 & 62 & 59 & 63 & 59 \\
\midrule
\multicolumn{8}{l}{\textit{Panel B: Large non-Chinese firms as comparison group}} \\[2pt]
Glasswing $\times$ Post & -0.614 & -1.241** & -0.871** & 0.237 & 0.949 & 0.120 & 0.122 \\
 & (0.541) & (0.506) & (0.390) & (0.830) & (0.725) & (0.186) & (0.383) \\[2pt]
\midrule
Pre-launch mean, Glasswing firms & 1.639 & 1.887 & 1.423 & 2.372 & 3.149 & 0.282 & 0.409 \\
Pre-launch mean, large non-Chinese firms & 1.057 & 1.151 & 0.660 & 1.774 & 1.943 & 0.374 & 0.696 \\
Last publication month & 2026-05 & 2026-05 & 2026-05 & 2026-05 & 2026-05 & 2026-05 & 2026-05 \\
Vulnerability-firm pairs & 16,395 & 16,395 & 16,395 & 16,395 & 16,395 & 16,393 & 16,393 \\
Clusters (firms) & 32 & 32 & 32 & 32 & 32 & 32 & 32 \\
\bottomrule
\end{tabular}}
\begin{minipage}{0.97\textwidth}\footnotesize \vspace{4pt} \textit{Notes:} Linear probability models on vulnerability-firm pairs with firm, publication-month and weakness-type fixed effects and severity controls; outcomes in percentage points. Panel A: Glasswing and Chinese firms, from October 2025 to the last publication month observed for the full window. Panel B: Glasswing firms and the 26 large non-Chinese firms, January 2024 to May 2026. Standard errors clustered by firm in parentheses. *, **, and *** denote significance at the 10, 5, and 1 percent levels.\end{minipage}
\end{table}

%% file: tables/tableA5_hunter_sources.tex
\begin{table}[!htbp]\centering
\caption{Sources of Finder Information}\label{tab:sources}
\small
\begin{tabular}{lrr}
\toprule
Source & Vulnerabilities covered & With acknowledgment (\%) \\
\midrule
CVE List (credits field) & 176,384 & 43.4 \\
Microsoft Security Update Guide & 21,615 & 30.3 \\
Chrome release notes & 3,324 & 100.0 \\
Cisco security advisories & 987 & 99.7 \\
Apple security notes & 2,412 & 96.9 \\
Android security bulletins & 934 & 100.0 \\
NVIDIA security bulletins & 421 & 100.0 \\
Adobe security bulletins & 2,521 & 94.0 \\
Oracle Critical Patch Updates & 6,387 & 9.3 \\
IBM security bulletins & 5,277 & 20.4 \\
Intel advisories (GitHub) & 580 & 97.9 \\
Qualcomm security bulletins & 645 & 47.3 \\
Fortinet PSIRT advisories & 481 & 92.7 \\
Dell security advisories & 574 & 3.7 \\
Huawei security bulletins & 529 & 3.2 \\
Samsung Mobile security updates & 383 & 89.8 \\
Red Hat security data & 620 & 1.3 \\
Siemens ProductCERT advisories & 630 & 0.0 \\
JPCERT/CC JVN advisories & 819 & 72.9 \\
Zero Day Initiative advisories & 3,730 & 100.0 \\
HackerOne disclosed reports & 375 & 100.0 \\
OSV.dev (open-source advisories) & 73,861 & 1.7 \\
GitHub Advisory Database & 118,458 & 6.2 \\
GitHub repository advisories (API) & 10,900 & 70.5 \\
Linux kernel fix commits (Reported-by) & 18,157 & 25.3 \\
Other fix commits on GitHub (Reported-by) & 38,890 & 16.5 \\
Reference links (write-ups, research teams) & 176,384 & 4.0 \\
Record's discovery field (internal or external) & 24,875 & 100.0 \\
\midrule
All sources combined & 176,384 & 73.0 \\
\bottomrule
\end{tabular}
\begin{minipage}{0.95\textwidth}\footnotesize \vspace{4pt} \textit{Notes:} Vulnerabilities published between January 2024 and October 2026. With acknowledgment: share of the covered vulnerabilities for which the source names at least one finder. The last row is the share of all sample vulnerabilities with an acknowledgment in at least one source.\end{minipage}
\end{table}

%% file: tables/tableA7_inferred_internal.tex
\begin{table}[!htbp]\centering
\caption{Finder Types with Inferred Internal Discovery}\label{tab:inferred}
\small
\resizebox{\textwidth}{!}{%
\begin{tabular}{lccccccc}
\toprule
\multicolumn{8}{l}{\textit{Vulnerabilities of treated and control firms}} \\[2pt]
 & \multicolumn{3}{c}{Full sample} & \multicolumn{4}{c}{Pre-treatment (Jan 2024--Mar 2026)} \\
\cmidrule(lr){2-4}\cmidrule(lr){5-8}
 & Obs. & Mean & SD & Treated firms & Non-treated firms & Difference & $t$-stat. \\
\midrule
Found by the vendor's internal team: explicit only & 176,384 & 0.040 & 0.195 & 0.133 & 0.014 & 0.118*** & 33.37 \\
Found by the vendor's internal team: including inferred & 176,384 & 0.147 & 0.354 & 0.247 & 0.094 & 0.153*** & 33.41 \\
Finder unknown: explicit only & 176,384 & 0.270 & 0.444 & 0.251 & 0.259 & -0.008* & -1.65 \\
Finder unknown: including inferred & 176,384 & 0.163 & 0.369 & 0.137 & 0.180 & -0.043*** & -11.22 \\
\bottomrule
\end{tabular}}
\begin{minipage}{0.97\textwidth}\footnotesize \vspace{4pt} \textit{Notes:} All vulnerabilities published between January 2024 and October 2026. Explicit only: the definition of the main tables. Including inferred: also counts as internal the vulnerabilities that the issuing vendor's advisory lists without credit and Linux kernel fixes whose code change credits no reporter. *, **, and *** denote significance at the 10, 5, and 1 percent levels.\end{minipage}
\end{table}

%% file: tables/tableA2_classification.tex
\begin{table}[!htbp]\centering
\caption{AI Exposure of Vulnerabilities}\label{tab:adcsum}
\small
\begin{tabular}{lrrrr}
\toprule
 & \multicolumn{2}{c}{All vulnerabilities} & Treated firms & Control firms \\ \cmidrule(lr){2-3}
Weakness type & Number & Share (\%) & Share (\%) & Share (\%) \\
\midrule
AI-exposed: memory safety & 27,841 & 17.3 & 36.3 & 19.8 \\
\quad Out-of-bounds write and buffer overflow & 12,156 & 7.6 & 13.9 & 8.4 \\
\quad Out-of-bounds read & 4,383 & 2.7 & 6.7 & 4.8 \\
\quad Use after free, double free, invalid free & 5,348 & 3.3 & 11.6 & 2.9 \\
\quad Null pointer dereference & 3,386 & 2.1 & 0.9 & 1.1 \\
\quad Integer overflow or wraparound & 1,752 & 1.1 & 2.5 & 1.5 \\
\quad Other memory-safety weaknesses & 1,758 & 1.1 & 2.4 & 1.2 \\
Other machine-verifiable & 54,163 & 33.7 & 13.5 & 17.5 \\
\quad Cross-site scripting & 22,502 & 14.0 & 3.5 & 11.8 \\
\quad SQL, command and code injection & 22,361 & 13.9 & 4.7 & 3.2 \\
\quad Path traversal & 5,804 & 3.6 & 2.6 & 1.4 \\
\quad Deserialisation, request forgery, XML entities & 5,251 & 3.3 & 2.8 & 1.3 \\
Other weakness types & 69,548 & 43.2 & 48.4 & 60.6 \\
No weakness type & 9,324 & 5.8 & 1.8 & 2.1 \\
\midrule
All vulnerabilities & 160,876 & 100.0 & N = 18,189 & N = 9,707 \\
\bottomrule
\end{tabular}
\begin{minipage}{0.95\textwidth}\footnotesize \vspace{4pt} \textit{Notes:} All vulnerabilities published between January 2024 and September 2026. Weakness types within a group can overlap. Treated and control firms: the 13 Glasswing firms and the 26 large non-Chinese firms.\end{minipage}
\end{table}

%% file: tables/tableA1_firms.tex
\begin{table}[!htbp]\centering
\caption{Firms in the Sample}\label{tab:firms}
\scriptsize
\resizebox*{!}{0.75\textheight}{%
\begin{tabular}{llrrrr}
\toprule
 & & \multicolumn{3}{c}{Vulnerabilities per month} & \\ \cmidrule(lr){3-5}
Group & Firm & Pre & Launch & Expansion & Total \\
\midrule
Glasswing & Microsoft & 99.6 & 177.5 & 592.2 & 5,414 \\
Glasswing & Google & 70.0 & 267.0 & 671.0 & 5,108 \\
Glasswing & IBM & 47.7 & 49.0 & 243.8 & 2,362 \\
Glasswing & Apple & 56.3 & 59.0 & 128.8 & 2,154 \\
Glasswing & Cisco & 27.8 & 29.5 & 74.8 & 1,109 \\
Glasswing & Broadcom & 12.1 & 20.5 & 71.5 & 654 \\
Glasswing & NVIDIA & 11.5 & 23.5 & 67.0 & 625 \\
Glasswing & Linux Foundation & 8.9 & 13.0 & 9.5 & 305 \\
Glasswing & Palo Alto Networks & 5.3 & 15.0 & 14.0 & 229 \\
Glasswing & Amazon (AWS) & 3.0 & 17.0 & 23.8 & 211 \\
Glasswing & Anthropic & 1.0 & 3.0 & 2.5 & 43 \\
Glasswing & CrowdStrike & 0.1 & 0.5 & 0.2 & 5 \\
\midrule
Chinese & Tenda & 41.5 & 49.0 & 29.8 & 1,337 \\
Chinese & TOTOLINK & 22.6 & 72.0 & 41.0 & 918 \\
Chinese & Huawei & 18.6 & 16.0 & 12.0 & 583 \\
Chinese & TP-Link & 6.7 & 10.0 & 18.0 & 274 \\
Chinese & Foxit & 6.9 & 8.0 & 15.5 & 263 \\
Chinese & Lenovo & 5.6 & 4.0 & 8.0 & 192 \\
Chinese & OpenHarmony & 5.0 & 5.0 & 0.0 & 144 \\
Chinese & WAVLINK & 4.4 & 7.0 & 2.2 & 143 \\
Chinese & Yonyou & 4.5 & 0.0 & 1.5 & 128 \\
Chinese & UTT & 3.0 & 6.0 & 4.0 & 110 \\
Chinese & Ruijie & 3.4 & 0.0 & 1.5 & 98 \\
Chinese & H3C & 2.4 & 2.5 & 2.8 & 81 \\
Chinese & ZTE & 1.7 & 7.0 & 4.2 & 76 \\
Chinese & Netcore & 0.8 & 0.0 & 11.0 & 65 \\
Chinese & Hikvision & 1.3 & 2.0 & 2.8 & 49 \\
Chinese & Dahua & 1.4 & 0.0 & 1.0 & 43 \\
Chinese & Tencent & 1.0 & 1.5 & 2.0 & 39 \\
Chinese & Baidu & 1.0 & 0.0 & 0.2 & 29 \\
Chinese & Unisoc & 0.5 & 0.0 & 2.2 & 23 \\
Chinese & Honor & 0.7 & 1.5 & 0.0 & 22 \\
Chinese & vivo & 0.6 & 0.0 & 1.0 & 21 \\
Chinese & Xiaomi & 0.7 & 0.0 & 0.0 & 18 \\
Chinese & DJI & 0.4 & 0.0 & 1.2 & 16 \\
Chinese & OPPO & 0.3 & 1.0 & 0.5 & 11 \\
Chinese & Alibaba & 0.1 & 0.5 & 1.0 & 8 \\
Chinese & Kingsoft & 0.2 & 0.0 & 0.0 & 6 \\
Chinese & ByteDance & 0.0 & 1.5 & 0.8 & 6 \\
Chinese & Other Chinese vendors (240 firms, sum) & 34.0 & 9.5 & 22.8 & 1,029 \\
\midrule
Other non-Chinese & Oracle & 27.9 & 67.0 & 718.5 & 3,762 \\
Other non-Chinese & Adobe & 64.1 & 52.5 & 143.8 & 2,412 \\
Other non-Chinese & Intel & 25.5 & 13.0 & 17.8 & 786 \\
Other non-Chinese & Qualcomm & 21.3 & 13.5 & 16.0 & 667 \\
Other non-Chinese & SAP & 15.3 & 16.5 & 20.5 & 528 \\
Other non-Chinese & Fortinet & 16.0 & 20.0 & 8.2 & 506 \\
Other non-Chinese & Juniper Networks & 9.3 & 13.5 & 5.5 & 300 \\
Other non-Chinese & AMD & 6.7 & 26.5 & 5.5 & 255 \\
Other non-Chinese & Arista Networks & 1.8 & 0.5 & 16.2 & 114 \\
Other non-Chinese & Rapid7 & 0.6 & 5.0 & 9.8 & 64 \\
Other non-Chinese & Tenable & 1.2 & 2.0 & 5.2 & 58 \\
Other non-Chinese & Meta & 1.1 & 2.5 & 2.5 & 45 \\
Other non-Chinese & Zscaler & 1.0 & 0.0 & 2.5 & 37 \\
Other non-Chinese & Salesforce & 1.3 & 0.5 & 0.0 & 35 \\
Other non-Chinese & Check Point & 0.6 & 3.0 & 2.8 & 34 \\
Other non-Chinese & Cloudflare & 1.0 & 0.0 & 1.2 & 31 \\
Other non-Chinese & Okta & 0.3 & 0.0 & 5.5 & 31 \\
Other non-Chinese & ServiceNow & 0.5 & 0.0 & 2.5 & 23 \\
Other non-Chinese & Marvell & 0.6 & 0.0 & 0.0 & 16 \\
Other non-Chinese & Qualys & 0.1 & 0.0 & 0.0 & 4 \\
\bottomrule
\end{tabular}}
\begin{minipage}{0.95\textwidth}\footnotesize \vspace{4pt} \textit{Notes:} Average number of vulnerabilities per month before the launch (January 2024 to March 2026), in April and May 2026, and from June to September 2026, and the total from January 2024 to September 2026. The row for other Chinese vendors sums over 240 vendors, each a firm in the regressions. Firms without any vulnerability in the period are not listed (JPMorganChase; Bank of America, Citigroup, Goldman Sachs, Morgan Stanley, SentinelOne and Wells Fargo) but remain in the regressions.\end{minipage}
\end{table}

%% file: tables/tableA_chinese_disclosure.tex
\begin{table}[!htbp]\centering
\caption{Chinese Disclosure before and after the Launch}\label{tab:chinese}
\small
\begin{tabular}{lccc}
\toprule
 & 2024 & Oct 2025 to & Apr 2026 to \\
 &  & Mar 2026 & Sep 2026 \\
\midrule
\multicolumn{4}{l}{\textit{A. Chinese vulnerabilities per month, by disclosure channel}} \\
All Chinese vulnerabilities & 180.1 & 157.2 & 192.7 \\
\quad Identifier issued by the Chinese vendor & 38.7 & 45.7 & 61.5 \\
\quad Identifier issued by an independent issuer & 108.8 & 99.2 & 123.2 \\
\qquad of which by VulDB & 37.6 & 70.2 & 74.2 \\
\quad Only in the Chinese national database & 32.6 & 12.3 & 8.0 \\
Share with a CVE identifier & 0.78 & 0.94 & 0.95 \\
Share issued by an independent issuer & 0.55 & 0.63 & 0.61 \\
Share of high or critical severity & 0.61 & 0.66 & 0.64 \\
\midrule
\multicolumn{4}{l}{\textit{B. Independent researchers on Chinese vulnerabilities, per month}} \\
Vulnerabilities credited to an independent researcher & 83.9 & 112.5 & 141.7 \\
\quad identifier issued by the Chinese vendor & 10.7 & 18.7 & 31.7 \\
\quad identifier issued by an independent issuer & 73.2 & 93.8 & 110.0 \\
Share credited to an independent researcher & 0.54 & 0.76 & 0.74 \\
Independent researchers credited & 28.8 & 49.5 & 57.0 \\
\quad credited on Chinese vulnerabilities for the first time &  & 34.7 & 44.3 \\
\midrule
\multicolumn{4}{l}{\textit{C. Timing (median days)}} \\
With a CVE identifier: reservation to publication & 7.9 & 1.6 & 8.1 \\
National database: submission to publication & 45 & 58 & 53 \\
\quad share already published with a CVE identifier & 0.56 & 0.88 & 0.93 \\
\quad entries & 907 & 652 & 758 \\
\bottomrule
\end{tabular}
\begin{minipage}{0.95\textwidth}\footnotesize \vspace{4pt} \textit{Notes:} Panels A and B report monthly means, Panel C medians. Panel A splits Chinese vulnerabilities by the issuer of the CVE identifier; shares are relative to all Chinese vulnerabilities in the month. Panel B counts independent researchers, excluding the vendor's own staff and researchers affiliated with a Glasswing firm; first credits are counted from January 2024. Panel C: days from reservation to publication in the CVE List, and from submission to publication in the China National Vulnerability Database.\end{minipage}
\end{table}

%% file: tables/tableA18_firm_loo.tex
\begin{table}[!htbp]\centering
\caption{Firm-Level Evidence: Leaving Out One Glasswing Firm at a Time}\label{tab:firmloo}
\small
\begin{tabular}{lcccc}
\toprule
Glasswing firm left out & Glasswing $\times$ Post & \shortstack{Pre-access\\ $N$ per month} & Firm-months & Clusters \\
\midrule
None (Table~\ref{tab:firm}, column 1) & 0.784*** (0.206) &  & 3,360 & 280 \\
\midrule
Microsoft & 0.740*** (0.218) & 101.3 & 3,348 & 279 \\
Google & 0.691*** (0.198) & 84.8 & 3,348 & 279 \\
Apple & 0.802*** (0.222) & 61.3 & 3,348 & 279 \\
IBM & 0.757*** (0.221) & 47.1 & 3,348 & 279 \\
Cisco & 0.789*** (0.223) & 25.8 & 3,348 & 279 \\
NVIDIA & 0.798*** (0.222) & 17.3 & 3,348 & 279 \\
Linux Foundation & 0.872*** (0.201) & 15.8 & 3,348 & 279 \\
Broadcom & 0.706*** (0.206) & 11.2 & 3,348 & 279 \\
Amazon (AWS) & 0.750*** (0.220) & 5.7 & 3,348 & 279 \\
Anthropic & 0.862*** (0.206) & 3.2 & 3,348 & 279 \\
Palo Alto Networks & 0.724*** (0.213) & 2.2 & 3,348 & 279 \\
CrowdStrike & 0.847*** (0.212) & 0.3 & 3,348 & 279 \\
JPMorganChase & 0.851*** (0.210) & 0.0 & 3,348 & 279 \\
\bottomrule
\end{tabular}
\begin{minipage}{0.92\textwidth}\footnotesize \vspace{4pt} \textit{Notes:} Table~\ref{tab:firm}, column (1), leaving out the Glasswing firm named in the row. Rows are ordered by the firm's vulnerabilities per month before access. Standard errors clustered by firm in parentheses. *, **, and *** denote significance at the 10, 5, and 1 percent levels.\end{minipage}
\end{table}

%% file: tables/tableR_inference.tex
\begin{table}[!htbp]\centering
\caption{Inference with Few Treated Firms}\label{tab:inference}
\small
\resizebox{\textwidth}{!}{%
\begin{tabular}{lcccc}
\toprule
Method & Estimate & s.e. & $p$-value & 95\% confidence set \\
\midrule
Firm-clustered (CR1), as in the paper & 0.784 & 0.206 & 0.0001 & [0.38, 1.19] \\
CR2 with Bell-McCaffrey degrees of freedom (13.2) & 0.784 & 0.204 & 0.0020 & [0.34, 1.22] \\
Wild cluster bootstrap, null imposed (Rademacher weights) & 0.784 &  & 0.0037 & [0.34, 1.24] \\
Wild cluster bootstrap, null imposed (Webb weights) & 0.784 &  & 0.0023 & [0.34, 1.20] \\
Ibragimov and M\"uller: t-test on 13 firm-by-firm estimates & 0.785 & 0.204 & 0.0023 & [0.34, 1.23] \\
Residual bootstrap without heteroskedasticity correction (Conley and Taber type) & 0.784 &  & 0.0000 & [0.64, 0.92] \\
Ferman and Pinto: A + B / M, M = 1 + mean monthly vulnerabilities before the launch (their form); variance estimated on all firms & 0.784 &  & 0.0000 & [0.40, 1.20] \\
Ferman and Pinto: A + B / M, M = 1 + mean monthly vulnerabilities before the launch (their form); variance estimated on Chinese firms only & 0.784 &  & 0.0000 & [0.44, 1.60] \\
Ferman and Pinto: A + B log(1 + M); variance estimated on all firms & 0.784 &  & 0.0000 & [-0.20, 1.08] \\
Ferman and Pinto: A + B log(1 + M); variance estimated on Chinese firms only & 0.784 &  & 0.0000 & [0.48, 1.26] \\
Ferman and Pinto: Size deciles; variance estimated on all firms & 0.784 &  & 0.0000 & [0.48, 1.18] \\
Ferman and Pinto: Size deciles; variance estimated on Chinese firms only & 0.784 &  & 0.0000 & [0.52, 1.22] \\
\bottomrule
\end{tabular}}
\begin{minipage}{0.97\textwidth}\footnotesize \vspace{4pt} \textit{Notes:} Table~\ref{tab:firm}, column (1). Confidence sets by test inversion on a grid with step 0.02. A bootstrap $p$-value of 0.0000 means that no draw exceeded the statistic. The Ferman-Pinto statistic is the cohort-weighted difference in pre-post means (launch partners from April 2026, IBM from May 2026).\end{minipage}
\end{table}

%% file: tables/tableA_opus.tex
\begin{table}[!htbp]\centering
\caption{Firm-Level Evidence: Access to Claude Opus 4.6}\label{tab:firmopus}
\small
\begin{tabular}{lcccc}
\toprule
 & (1) & (2) & (3) & (4) \\
 & \shortstack{Baseline} & \shortstack{Poisson} & \shortstack{Excluding\\ Anthropic} & \shortstack{Glasswing\\ trend} \\
\midrule
Glasswing $\times$ Post (Mythos Preview) & 0.486** & 1.035*** & 0.589*** & 0.438* \\
 & (0.235) & (0.343) & (0.228) & (0.238) \\[2pt]
Glasswing $\times$ Post Opus 4.6 & 0.452*** & 0.179 & 0.415** & 0.373** \\
 & (0.161) & (0.240) & (0.169) & (0.147) \\[2pt]
Glasswing $\times$ trend (months) &  &  &  & 0.0126* \\
 &  &  &  & (0.0071) \\[2pt]
\midrule
Dependent variable & $\log(1+N)$ & $N$ & $\log(1+N)$ & $\log(1+N)$ \\
Estimator & OLS & Poisson & OLS & OLS \\
Window start & Oct 2025 & Oct 2025 & Oct 2025 & Jan 2024 \\
Firm-months & 3,360 & 1,188 & 3,348 & 9,240 \\
Clusters (firms) & 280 & 99 & 279 & 280 \\
\bottomrule
\end{tabular}
\begin{minipage}{0.92\textwidth}\footnotesize \vspace{4pt} \textit{Notes:} Table~\ref{tab:firm}, columns (1) to (4), adding Glasswing $\times$ Post Opus 4.6, which equals one for the Glasswing firms from February 2026. Standard errors clustered by firm in parentheses. *, **, and *** denote significance at the 10, 5, and 1 percent levels.\end{minipage}
\end{table}

%% file: tables/tableA11_firm_gap.tex
\begin{table}[!htbp]\centering
\caption{The Disclosure Gap: New Discoveries or a Backlog?}\label{tab:firmgap}
\small
\resizebox{\textwidth}{!}{%
\begin{tabular}{lcccccc}
\toprule
 & \multicolumn{4}{c}{Firm-month, $\log(1+N)$} & \multicolumn{2}{c}{Vulnerability level} \\
\cmidrule(lr){2-5}\cmidrule(lr){6-7}
 & (1) & (2) & (3) & (4) & (5) & (6) \\
 & \shortstack{$N$: gap\\ $\leq$ 7 days} & \shortstack{$N$: gap\\ 8--90 days} & \shortstack{$N$: gap\\ $>$ 90 days} & \shortstack{$N$: no CVE\\ identifier} & \shortstack{Gap $\leq$ 7 days\\ (\%)} & \shortstack{$\log(1+\text{gap})$} \\
\midrule
Glasswing $\times$ Post & 0.524** & 0.941*** & 0.063 & -0.004 & 27.745** & -1.422*** \\
 & (0.257) & (0.298) & (0.180) & (0.005) & (11.914) & (0.483) \\[2pt]
\midrule
Pre-launch mean, Glasswing firms & 3.291 & 11.861 & 14.013 & 0.000 & 11.285 & 4.114 \\
Pre-launch mean, Chinese firms & 0.342 & 0.156 & 0.044 & 0.046 & 63.061 & 1.801 \\
Firm FE & Yes & Yes & Yes & Yes & Yes & Yes \\
Year-month FE & Yes & Yes & Yes & Yes & Yes & Yes \\
Observations & 3,360 & 3,360 & 3,360 & 3,360 & 13,177 & 13,177 \\
Clusters (firms) & 280 & 280 & 280 & 280 & 63 & 63 \\
\bottomrule
\end{tabular}}
\begin{minipage}{0.97\textwidth}\footnotesize \vspace{4pt} \textit{Notes:} Columns (1) to (4): equation~\eqref{eq:firm} with $N$ the number of the firm's vulnerabilities with a disclosure gap in the range of the column heading; column (4) counts entries without a CVE identifier, so the four counts add up to the total in Table~\ref{tab:firm}. Columns (5) and (6): vulnerability-firm pairs with a CVE identifier. Standard errors clustered by firm in parentheses. *, **, and *** denote significance at the 10, 5, and 1 percent levels.\end{minipage}
\end{table}

%% file: tables/tableA14_firm_finder.tex
\begin{table}[!htbp]\centering
\caption{Firm-Level Evidence by Finder Type}\label{tab:firmfinder}
\small
\resizebox{\textwidth}{!}{%
\begin{tabular}{lcccccccc}
\toprule
 & & & \multicolumn{3}{c}{External} & & \multicolumn{2}{c}{Including inferred} \\
\cmidrule(lr){4-6}\cmidrule(lr){8-9}
 & (1) & (2) & (3) & (4) & (5) & (6) & (7) & (8) \\
 & \shortstack{All} & \shortstack{Internal\\ team} & \shortstack{All external} & \shortstack{Other\\ organization} & \shortstack{Individual} & \shortstack{Unknown} & \shortstack{Internal} & \shortstack{Unknown} \\
\midrule
Glasswing $\times$ Post & 0.784*** & 0.576** & 0.650*** & 0.587*** & 0.559*** & 0.580*** & 0.603** & 0.464** \\
 & (0.206) & (0.272) & (0.161) & (0.157) & (0.143) & (0.213) & (0.264) & (0.209) \\[2pt]
\midrule
Pre-launch mean of $N$, Glasswing firms & 29.165 & 3.949 & 17.899 & 7.038 & 10.861 & 7.316 & 7.582 & 3.684 \\
Post-launch mean of $N$, Glasswing firms & 115.532 & 31.870 & 54.000 & 21.195 & 32.805 & 29.662 & 43.026 & 18.506 \\
Pre-launch mean of $N$, Chinese firms & 0.589 & 0.000 & 0.476 & 0.069 & 0.407 & 0.113 & 0.066 & 0.047 \\
Firm FE & Yes & Yes & Yes & Yes & Yes & Yes & Yes & Yes \\
Year-month FE & Yes & Yes & Yes & Yes & Yes & Yes & Yes & Yes \\
Firm-months & 3,360 & 3,360 & 3,360 & 3,360 & 3,360 & 3,360 & 3,360 & 3,360 \\
Clusters (firms) & 280 & 280 & 280 & 280 & 280 & 280 & 280 & 280 \\
\bottomrule
\end{tabular}}
\begin{minipage}{0.97\textwidth}\footnotesize \vspace{4pt} \textit{Notes:} Equation~\eqref{eq:firm} with $N$ the number of the firm's vulnerabilities found by the finder type in the column. Columns (2), (3) and (6) add up to column (1). Columns (7) and (8) also count as internal the vulnerabilities that the vendor's advisory lists without credit and Linux kernel fixes without a credited reporter. Standard errors clustered by firm in parentheses. *, **, and *** denote significance at the 10, 5, and 1 percent levels.\end{minipage}
\end{table}

%% file: tables/table10_exploited_counts.tex
\begin{table}[!htbp]\centering
\caption{Number of Exploited Vulnerabilities and Exploits}\label{tab:explcount}
\small
\resizebox{\textwidth}{!}{%
\begin{tabular}{lcccccc}
\toprule
 & (1) & (2) & (3) & (4) & (5) & (6) \\
 & \shortstack{Exploited\\ 1 month} & \shortstack{Exploited\\ 2 months} & \shortstack{Exploited\\ 3 months} & \shortstack{Zero-day} & \shortstack{N-day\\ 3 months} & \shortstack{Exploit code\\ 3 months} \\
\midrule
Glasswing $\times$ Post & 0.100 & -0.225 & -0.413 & -0.347 & -0.225 & 0.503* \\
 & (0.445) & (0.363) & (0.375) & (0.488) & (0.659) & (0.266) \\[2pt]
\midrule
Firm and month FE & Yes & Yes & Yes & Yes & Yes & Yes \\
Mean of dependent variable & 0.198 & 0.220 & 0.226 & 0.157 & 0.069 & 0.403 \\
Observations & 493 & 522 & 522 & 406 & 464 & 667 \\
Clusters (firms) & 17 & 18 & 18 & 14 & 16 & 23 \\
\bottomrule
\end{tabular}}
\begin{minipage}{0.95\textwidth}\footnotesize \vspace{4pt} \textit{Notes:} Firm-month Poisson regressions for the Glasswing firms and the 26 large non-Chinese firms, January 2024 to May 2026. The dependent variable is the number of the firm's vulnerabilities published in the month that meet the condition in the column heading (Table~\ref{tab:expl}). Standard errors clustered by firm in parentheses. *, **, and *** denote significance at the 10, 5, and 1 percent levels.\end{minipage}
\end{table}